\documentstyle[preprint,revtex,eqsecnum]{aps}
\tightenlines
\begin{document}
\draft
\begin{title}
Scaling theory of the Mott-Hubbard
metal-insulator\\ transition in one dimension
\end{title}
\author{C.~A.~Stafford} 
\begin{instit}
Department of Physics, University of Maryland, College Park, Maryland 20742
\end{instit}
\author{A.~J.~Millis} 
\begin{instit}
AT\&T Bell Laboratories, Murray Hill, New Jersey 07974
\end{instit}
\receipt{27 January 1993}
\begin{abstract}
We use the Bethe ansatz equations to calculate the charge stiffness
$D_{\rm c} = (L/2) d^2 E_0/d\Phi_{\rm c}^2|_{\Phi_{\rm c}=0}$
of the one-dimensional repulsive-interaction Hubbard
model for electron densities close to the Mott insulating value of one
electron per site ($n=1$),
where $E_0$ is the ground state energy, $L$
is the circumference of the system (assumed to have periodic boundary
conditions), and
$(\hbar c/e)\Phi_{\rm c}$ is the magnetic
flux enclosed.
We obtain an exact result for the asymptotic form of $D_{\rm c}(L)$
as $L\rightarrow \infty$ at $n=1$, which defines and yields an analytic
expression for
the correlation length $\xi$ in the
Mott insulating phase of the model as a
function of the on-site repulsion $U$.
In the vicinity of the zero temperature 
critical point $U=0$, $n=1$, we show that
the charge stiffness has the hyperscaling form
$D_{\rm c}(n,L,U)=Y_+(\xi \delta, \xi/L)$, where $\delta =|1-n|$
and $Y_+$ is a universal scaling function which we calculate.
The physical significance of $\xi$
in the metallic phase of the model is that it
defines the characteristic size of the charge-carrying
solitons, or {\em holons}.  We
construct an explicit mapping for arbitrary $U$ and $\xi \delta \ll 1$
of the holons onto weakly interacting spinless fermions, and
use this mapping to obtain
an asymptotically
exact expression for the
low temperature thermopower near the metal-insulator transition, which is a
generalization to arbitrary $U$ of a result previously obtained using a
weak-coupling approximation, and implies hole-like transport
for $0<1-n\ll\xi^{-1}$.
\end{abstract}

\narrowtext
\section{INTRODUCTION}
\label{intro}

The Mott-Hubbard metal-insulator transition \cite{mott}
is one of the fundamental problems
in electronic condensed matter physics.  The assertion by Anderson \cite{phil}
that
the high-$T_{\rm c}$ $\mbox{Cu} \mbox{O}_2$ superconductors should be viewed
as lightly doped Mott insulators has stimulated much work on the
problem.  In this paper, we use the Bethe ansatz equations \cite{lwu,ss}
to calculate the zero temperature
conductivity and low temperature thermopower of
the one-dimensional (1D) repulsive-interaction Hubbard model for
electron densities close to the Mott insulating value of one electron
per site.

One way to characterize interacting fermion systems is via the
zero temperature frequency dependent conductivity $\tilde{\sigma} (\omega)$,
which is the linear response of the system to a spatially uniform,
time dependent electric field.  In general, $\tilde{\sigma}(\omega)$ has the
form
\begin{equation}
\tilde{\sigma}(\omega)=2 D_{\rm c} \left(\frac{i}{\omega}+\pi
\delta(\omega)\right) + \tilde{\sigma}_{reg}(\omega),
\label{cond}
\end{equation}
where ${\displaystyle \lim_{\omega \rightarrow 0}}\,\omega
\tilde{\sigma}_{reg} (\omega)=0$.
The coefficient $D_{\rm c}$ is the charge stiffness.  In a Galilean-invariant
system with $n$ electrons per unit volume of charge $e$ and mass $m$
interacting via velocity-independent forces, $D_{\rm c}
=ne^2/2m$ and $\tilde{\sigma}_{reg}(\omega)=0$,
independent of the interactions.
For electrons moving in an
external potential, $D_{\rm c}$ may depend on the interactions and on the
potential.  Walter Kohn has argued \cite{kohn}
that for  spatially infinite systems with a discrete translational
invariance
two cases are possible:  either $D_{\rm c}=0$ and $\displaystyle{\lim_{\omega
\rightarrow 0}}\,{\mbox Re}[\tilde{\sigma}_{reg}(\omega)]=0$,
or $D_{\rm c} > 0$.
If $D_{\rm c} =0$, the
material is insulating; if $D_{\rm c} >0$, the material has the infinite dc
conductivity and perfect diamagnetism expected of a perfect metal.
$D_{\rm c}$ is thus an order parameter \cite{kohn}
for the Mott-Hubbard metal-insulator transition
which may occur in a nondisordered system as the electron concentration
and interaction strength are varied.

In a system of finite length $L$ with periodic boundary conditions,
$D_{\rm c}$ is in general
non-vanishing \cite{persistcurrent}.
In a previous paper \cite{ours}, we showed numerically and via a
large-$U$ approximation that in the Mott insulating
phase of the 1D Hubbard model, which occurs at $n=1$ for repulsive
interactions, the large-$L$ behavior of $D_{\rm c}$ is
\begin{equation}
D_{\rm c} (L) \sim L^{1/2}\exp(-L/\xi).
\label{eq1.2}
\end{equation}
This equation defines the correlation length $\xi$ of the Mott
insulator, which is a function of the on-site repulsion $U$.
In this paper, we derive Eq.~(\ref{eq1.2}) analytically for arbitrary
$U$, and show that the correlation length so defined
is important also in the metallic phase of the model.
In particular, we show that in the vicinity of the zero temperature
critical point $U=0$, $n=1$, the charge stiffness assumes the
hyperscaling form:
\begin{equation}
D_{\rm c}(n,L,U)= Y_+(\xi \delta,\xi/L),
\label{eq1.3}
\end{equation}
where $\delta=|1-n|$ and $Y_+$ is a (presumably universal
\cite{kimweichman}) scaling function which we calculate.

The physical significance of $\xi$
in the metallic phase of the model is that it
defines the characteristic size of the charge-carrying
solitons, or {\em holons}.
We show that Woynarovich's reformulation \cite{woy4} of the Bethe ansatz
equations of the 1D Hubbard model in
terms of the parameters of the charge
excitations only is formally equivalent to
an asymptotic (large $L$) Bethe ansatz for {\em holons},
and use this asymptotic Bethe ansatz to construct
an explicit mapping for arbitrary $U$ and $\xi \delta \ll 1$
of the low-lying charge degrees of freedom of the
model onto weakly interacting spinless fermions.
Using this mapping, we obtain asymptotically exact expressions
for the charge stiffness and low temperature thermopower near the
metal-insulator transition.
Our result for the thermopower extends and makes more rigorous previous
work of Schulz \cite{schulztherm} which was
based on a weak-coupling approximation, and
implies that the transport is hole-like for $0<1-n\ll\xi^{-1}$.

The paper is organized as follows:  In Sec.~\ref{form}, we review the
linear response arguments and Bethe ansatz equations used to calculate
$D_{\rm c}$ in 1D.  In Sec.~\ref{localz}, we present an exact result for
the charge stiffness at $n=1$ in the large-$L$ limit, verifying
Eq.~(\ref{eq1.2}), and show that the correlation length so defined
also governs the exponential decay of the equal-time
single-particle Green's function at $n=1$ in
both the weak- and strong-coupling
limits, as well as the pairing correlations in the ground state of the
attractive 1D Hubbard model at $n=1$.  In Sec.~\ref{criticalscaling},
we determine the scaling behavior of
$D_{\rm c}$ near the metal-insulator critical point.  The
total optical spectral weight $\pi N_{{\rm tot}}
\equiv \int_{0}^{\infty} \mbox{Re}\,[\tilde{\sigma}(\omega)] d \omega$
is also discussed.  In Sec.~\ref{finite}, we calculate the finite-size
corrections to $D_{\rm c}$ and $N_{\rm tot}$ in the regime $L>\xi$, and
discuss the implications for the interpretation of
numerical calculations of $\tilde{\sigma}(\omega)$ in interacting
fermion systems on small
clusters.  In  Sec.~\ref{sec.holonband}, we determine the charge
excitations, charge stiffness,
and low temperature thermopower near the metal-insulator
transition.
We summarize our results in Sec.~\ref{conclusions},
and comment briefly on their relevance for
the physics of high-$T_{\rm c}$ superconducting materials.
The details of the calculations of $D_{\rm c}$
at and near $n=1$ in the large-$L$ limit are given in
Appendixes \ref{calcxi} and \ref{appB}, respectively.

\section{FORMALISM}
\label{form}

We consider
the Hubbard model of spin-1/2 fermions hopping with matrix element $t$
between nearest-neighbor sites of a one-dimensional lattice with unit
lattice constant, and subject to a repulsive interaction $U$ when two
fermions (of opposite spin) occupy the same lattice site.
We impose periodic boundary conditions,
and thread the system with a (dimensionless)
spin-dependent flux $\Phi_{\sigma}$, which we represent by
a uniform spin-dependent vector potential $ A_{\sigma} =
(\hbar c/e)\Phi_{\sigma}/L$, where $L$ is the circumference of the
system.  The Hamiltonian may be written
\FL
\begin{equation}
 H 
 = -t \sum_{l,\sigma}
\left[e^{i\Phi_{\sigma}/L} c_{l+1\sigma}^{\dagger}
c_{l\sigma} + H.c. \right]
 + U \sum_{l} n_{l\uparrow} n_{l\downarrow}, \label{hubham}
\end{equation}
where $n_{l\sigma}=c_{l\sigma}^{\dagger}c_{l\sigma}$ and
the sums run from $l=1$ to $L$ and $\sigma = \uparrow, \downarrow$.
In the following, we set $e=\hbar=t=1$ unless explicitly stated.

The numbers $M$ and $M'$ of up- and down-spin electrons are constants of the
motion; we denote the minimum energy eigenvalue in a given sector ($M,M'$)
by $E(M,M',U,\Phi_{\uparrow},\Phi_{\downarrow})$.  The sign of the interaction
$U$ in Eq.~(\ref{hubham}) can be reversed by a particle-hole transformation
on the down-spin electrons \cite{robasz},
\begin{equation}
\begin{array}{ccc}
c_{l\downarrow}^{\dagger} \rightarrow (-1)^{l} c_{l\downarrow},&\;\;\;\;&
c_{l\uparrow}^{\dagger} \rightarrow  c_{l\uparrow}^{\dagger}, \\
c_{l\downarrow} \rightarrow (-1)^{l} c_{l\downarrow}^{\dagger},&\;\;\;\;&
c_{l\uparrow} \rightarrow  c_{l\uparrow},
\end{array}
\label{e-h}
\end{equation}
leading to the following relation,
\FL
\begin{equation}
E(M,M',U,\Phi_{\uparrow},\Phi_{\downarrow}) =
E(M,L-M',-U,\Phi_{\uparrow},-\Phi_{\downarrow}) + MU,
\label{negu}
\end{equation}
which we will exploit in Secs.~\ref{localz} and \ref{conclusions}.

To simplify matters, we consider only systems where
the total number of electrons
$N=M+M'$ is even.  Then the ground state
is a singlet \cite{triplet} and the ground state energy is
$E_{0} = E(N/2,N/2,U,\Phi_{\uparrow},\Phi_{\downarrow})$.
The charge stiffness $D_{\rm c}$ is
defined by \cite{kohn,ss}
\begin{equation}
D_{\rm c} = \left. \frac{1}{2} \frac{d^{2} (E_0/L)}{d (\Phi_{\rm c}/L)^{2}}
\right|_{\Phi_{\uparrow}=\Phi_{\downarrow} = 0},
\label{stiff}
\end{equation}
where $\Phi_{\rm c} = (\Phi_{\uparrow}+\Phi_{\downarrow})/2$.
Applying second-order perturbation theory in $\Phi_{\rm c}$
to Eq.~(\ref{hubham}) gives \cite{ss,mc}
\begin{equation}
D_{\rm c} = \frac{1}{L}
\left( \frac{1}{2}\langle -T \rangle -\sum_{\nu \neq 0}
\frac{|\langle \nu| J_{p} |0\rangle|^{2}}{E_{\nu}-E_0}\right),
\label{stiffpert}
\end{equation}
where $T = - \sum (c_{l+1 \sigma}^{\dagger} c_{l\sigma} + H.c.)$
is the kinetic energy operator, $ J_{p} = -i\sum (c_{l+1\sigma}^{\dagger}
c_{l\sigma} - H.c.)$ is the paramagnetic current operator,
$\langle\mbox{ } \rangle$ denotes the expectation value
in the ground state, and all quantities are evaluated at $\Phi_{\uparrow}=
\Phi_{\downarrow}=0$.

We next consider threading the Hubbard ring with
a time-dependent flux $\Phi_{\uparrow}(t) =
\Phi_{\downarrow}(t) = \Phi_{\rm c}(0) \exp(-i\omega t)$,
which leads to a uniform time-dependent electric field by Faraday's law.
Standard
linear-response arguments applied to Eq.~(\ref{hubham}) give the
frequency-dependent conductivity at zero temperature \cite{kohn,ss,mc},
\FL
\begin{equation}
\tilde{\sigma}(\omega) = \frac{i}{\omega+i0^+} \left(\frac{\langle
-T\rangle}{L} + \frac{2}{L} \sum_{\nu \neq 0}
\frac{|\langle \nu| J_{p}
|0\rangle|^{2}}{\omega -E_{\nu}+E_{0}+i0^+}\right).
\label{cond2}
\end{equation}
The conductivity thus has the form (\ref{cond}) with $D_{\rm c}$ given by
Eq.~(\ref{stiffpert}).
The high-frequency behavior of $\mbox{Im}\,[\tilde{\sigma}(\omega)]$
leads, via the Kramers-Kronig relations which link the real and
imaginary parts of $\tilde{\sigma}(\omega)$, to an {\em f}-sum rule for
the total optical spectral weight \cite{kohn,mald},
\begin{equation}
\pi N_{{\rm tot}} \equiv
\int_{0}^{\infty} \mbox{Re}\,[\tilde{\sigma}(\omega)] d \omega
= \frac{\pi}{2L}\langle -T \rangle. \label{total}
\end{equation}
The Hellman-Feynman theorem gives
$\langle T\rangle = E_0 -U\partial E_0/\partial
U$.  Thus both $N_{{\rm tot}}$ and $D_{\rm c}$ may be obtained as
derivatives of the ground state energy.

The energy and momentum of the eigenstates of Eq.~(\ref{hubham})
can be expressed via a generalization of Bethe's ansatz as \cite{lwu,ss}
\begin{equation}
E = -2 \sum_{n=1}^{N} \cos k_{n},
\label{ener}
\end{equation}
\begin{equation}
P = \sum_{n=1}^{N} k_{n}, 
\label{ptot}
\end{equation}
where the {\em pseudomomenta} $k_n$ are a set of distinct numbers
determined by \cite{ss}
\FL
\begin{equation}
L k_{n}  =  2\pi I_{n} + \Phi_{\downarrow} -  \sum_{\alpha=1}^{M} 2\tan^{-1}
\frac{\sin k_{n} - \Lambda_{\alpha}}{U/4},
\label{keqn}
\end{equation}
and the {\em spin rapidities} $\Lambda_{\alpha}$ are
a set of distinct numbers related to the $k_n$ by \cite{ss}
\widetext
\begin{equation}
 \sum_{n=1}^{N} 2\tan^{-1} \frac{\Lambda_{\alpha} -
 \sin k_{n}}{U/4}  =  2 \pi
J_{\alpha} +\Phi_{\uparrow}-\Phi_{\downarrow}
+  \sum_{\beta = 1}^{M} 2\tan^{-1}
\frac{\Lambda_{\alpha} -\Lambda_{\beta}}{U/2}.
\label{leqn}
\end{equation}
\narrowtext
Here $\{I_n; n=1,\ldots, N\}$ and $\{J_{\alpha}; \alpha=1,\ldots, M\}$
are the quantum numbers which specify the state of the system
\cite{complete}:  the $I_n$ describe the charge degrees of freedom
\cite{ov,cfcoll,woy4}, and
are distinct integers (half-integers) for $M$ even (odd) \cite{lwu}, while the
$J_{\alpha}$ describe the spin degrees of freedom \cite{ov,cfcoll,woy4},
and are integers (half-integers) for $N-M$ odd (even) \cite{lwu}.
For the ground state, the quantum
numbers are consecutive integers (or half-integers) centered
about the origin \cite{lwu}.

In the following, we solve the Bethe ansatz equations
(\ref{keqn}) and (\ref{leqn})
analytically and numerically in various limits to obtain $D_{\rm c}$ and
$N_{{\rm tot}}$
as a function of $n$, $L$, and $U$.  The knowledge of the
eigenstates of $H$ also allows us to calculate the low temperature
thermopower near half filling.

\section{THE CORRELATION LENGTH OF THE MOTT INSULATOR}
\label{localz}

\subsection{Definition of $\xi$}

The 1D Hubbard model for $n=1$ and $U>0$ is known to have an insulating
ground state \cite{lwu}, so that the charge stiffness $D_{\rm c}$ vanishes for
an infinite system.  In a previous paper \cite{ours}, we showed
numerically and via a large-$U$ approximation that for a large finite ring of
circumference $L$ at $n=1$
\FL
\begin{equation}
D_{\rm c}(L)=(-1)^{\frac{L}{2}+1} L^{1/2} D(U) \exp[-L/\xi(U)];\;\;\;\;\;
L \rightarrow \infty
\label{stiffans}
\end{equation}
($L$ even), where $D(U)$ is a positive $U$-dependent number.
Eq.~(\ref{stiffans}) serves to define the correlation length $\xi(U)$ in the
Mott insulating phase of the 1D Hubbard model.
We have subsequently derived
Eq.~(\ref{stiffans}) analytically for arbitrary $U$.  The details are
given in Appendix~\ref{calcxi}.  The result for the correlation length
$\xi(U)$ is
\begin{equation}
1/\xi(U) =
 \frac{4}{U} \int_{1}^{\infty} d y \, \frac{\ln (y +
\sqrt{y^{2} - 1})}{\cosh (2 \pi y/U)}.
\label{xinv}
\end{equation}
The result for $D(U)$
is given in Eq.~(\ref{eq.coeff}).
The limiting behavior of $D(U)$ is:
\begin{equation}
D(U) =  \left\{\begin{array}{ll}
(2/\pi \xi)^{1/2}\;\;\;\; &
U \rightarrow 0, \\
A\,U &
U \rightarrow \infty,
\end{array}
\right.
\label{limitcoeff}
\end{equation}
where $A \simeq 0.147376$.

The negative sign of $D_{\rm c}$ when $L$ is a multiple of 4 indicates
that the persistent current of the ring is paramagnetic, as discussed
in Refs.~\cite{ours,fye}.  Orbital paramagnetism is a generic feature of
half-filled single-band $4n$ electron systems, and was first observed many
years ago \cite{nmr} in NMR spectra of [16]annulene and [24]annulene,
larger analogs of benzene.
The full flux dependence of the ground state energy of the Mott insulator is
\FL
\begin{equation}
E_0 (\Phi_{\rm c}) -E_0 (0) = \frac{2 D_{\rm c} (L)}{L}
(1-\cos \Phi_{\rm c}); \;\;\;\;\; L \rightarrow \infty.
\label{eperiodic}
\end{equation}
By contrast, in the metallic phase at $n\neq 1$, the flux dependence of
the ground state energy in the large-$L$ limit
is $E_0(\Phi_{\rm c})-E_0(0) = D_{\rm
c}\Phi_{\rm c}^2/L$ for $|\Phi_{\rm c}|\ll 2\pi L$, and the periodicity
$E_0(\Phi_{\rm c})=E_0(\Phi_{\rm c}+2\pi)$
required
by gauge invariance \cite{byersyang} is restored by
level crossings.

\subsection{Relation of $\xi$ to the Single-Particle Green's Function}
\label{greenxi}

The zero temperature equal-time single-particle Green's
function is defined as
\begin{equation}
G_{\sigma \sigma'}(|x-x'|) =
 \langle 0| c_{x'\sigma'}^{\dagger} c_{x\sigma} +
c_{x\sigma}^{\dagger} c_{x'\sigma'} | 0 \rangle.
\label{greenf}
\end{equation}
 From Eq.~(\ref{hubham}) it follows that $G_{\sigma \sigma'}(|x-x'|)=
G_{\sigma \sigma}(|x-x'|) \delta_{\sigma, \sigma'}$.
We wish to show that
the Green's function of the Mott insulator
has the form (at $L=\infty$)
\begin{equation}
G_{\sigma \sigma}(|x|)\sim \exp(-|x|/\xi)\;\;\; \mbox{as} \;\;\;
|x|\rightarrow \infty.
\label{corr}
\end{equation}
In order to do so, it is useful first to consider some limiting cases.

The weak-coupling limit of $\xi$ may be obtained from an asymptotic expansion
of Eq.~(\ref{xinv}) around $U=0$.  We obtain
\widetext
\begin{equation}
\lim_{U \rightarrow 0} \xi = \frac{\pi}{2} (t/U)^{1/2}
e^{2\pi t/U} (1 + U/16 \pi t + \cdots )
= \frac{2t + U/2\pi + \cdots}{\Delta (U,t)},
\label{smallU}
\end{equation}
\narrowtext
where we have made the hopping matrix element $t$ explicit, and
\begin{equation}
2 \Delta(U,t) =
\frac{16 t^{2}}{U} \int_{1}^{\infty} \frac{(y^{2} - 1)^{1/2}
d y}{\sinh (2 \pi t y/U)}
\label{gap}
\end{equation}
is the Lieb-Wu charge gap \cite{lwu,woy4,ov}.
The low energy (continuum limit) physics
of the 1D Hubbard model has been studied extensively by
bosonization techniques \cite{bose,haldanellt,bose2,schulzcond,giamarchi}.
In the low energy sector, the effective Hamiltonian
separates into independent terms describing the charge and spin degrees
of freedom \cite{bose,haldanellt,bose2,schulzcond,giamarchi}.
Near $n=1$ and in the weak-coupling limit, the Hamiltonian
for the charge degrees of freedom is of the relativistic
sine-Gordon form \cite{bose,giamarchi}.
The quantity which plays the role of the speed of light $c$ in the
Hamiltonian is
given to leading order in $U$ at $n=1$ by \cite{bose,giamarchi}
\begin{equation}
c = 2t + U/2\pi + \cdots.         \label{light}
\end{equation}
The elementary objects in the sine-Gordon model are solitons
\cite{coleman} carrying charge $e$ and
obeying the relativistic dispersion
relation $E_{k} = (c^{2} k^{2} + \Delta^{2})^{1/2}$ \cite{vjemery,schulz2}.
The {\em quantum soliton
length} $\xi_{s} =c/\Delta$ defines the characteristic size of the solitons
\cite{haldane1}, so we see from Eqs.~(\ref{smallU}) and (\ref{light})
that in the weak-coupling limit the correlation
length is equal to the quantum soliton length.

The mapping onto the weak-coupling Hubbard model fixes the number
density of solitons as $n_s=1-n$ \cite{giamarchi}.
The operator $c_{x\sigma}$ involves creating at least one soliton
(plus spin excitations).
Thus, if we evaluate Eq.~(\ref{greenf}) at a time-like separation at
$n=1$, we
find $G_{\sigma \sigma}
(t-t') \sim \exp [-i \Delta (t-t')] \;+$ terms involving
higher energies.
Using the Lorentz invariance of the weak-coupling field theory
to rotate back to a space-like separation, we obtain $G_{\sigma
\sigma}(|x-x'|)
\sim \exp (- \Delta |x-x'|/c)$ \cite{haldane2}.
The gapless spin degrees of freedom
correct this expression only by a power of $|x-x'|$.
We have thus established Eq.~(\ref{corr}) in the
weak-coupling limit.

Now let us consider the strong-coupling limit.
 From Eq.~(\ref{xinv}) it follows \cite{xinv2} that
\begin{equation}
\xi^{-1} = \ln (U/at);\;\;\;\;\; U \rightarrow \infty,
\label{largeU}
\end{equation}
where $a = (\Gamma (1/4)/\sqrt{2 \pi})^{4} \simeq 4.377$ (this result was
obtained previously by us in Ref.~\cite{ours}).
Eqs.~(\ref{stiffans}) and (\ref{limitcoeff}) then imply that
$D_{\rm c}(L)= A U L^{1/2} (at/U)^{L}$, where $A
\simeq 0.147376$.
The correlation length may thus be obtained in the large-$U$ limit from an
expansion of the ground state energy in powers of $t/U$.
The leading contribution
to $D_{\rm c}(L)$ comes from the lowest order
term which is flux dependent, and is \cite{ours}
\widetext
\begin{equation}
\lim_{U\rightarrow\infty} \frac{2D_{\rm c}(L)}{L}
= \frac{t^{L}}{U^{L-1}} \left[ \sum_{P \in {\cal S}_{L}}
\langle 0_{\infty}| v_{P_{L}}^{\dagger} \prod_{i=1}^{L-1}\left( {\bf D}^{-1}
v_{P_{i}}^{\dagger}\right) |0_{\infty}\rangle + H.c. \right],
\label{bw}
\end{equation}
\narrowtext
where $ v_{i}^{\dagger} = \sum_{\sigma} c_{i+1 \sigma}^{\dagger} c_{i\sigma}$,
${\bf D}^{-1} = 1/ \sum_{i=1}^{L} n_{i\uparrow} n_{i\downarrow}$, $|0_{\infty}
\rangle = \lim_{U \rightarrow \infty} |0 \rangle$, and ${\cal S}_{L}$ is the
group of permutations of $L$ objects.
This expression for $D_{\rm c}$ is a sum over
all processes which transport one unit of charge
around the ring in the minimum number of steps, $L$.
In the large-$U$ limit, the Green's function
$G_{\sigma \sigma}(L-1)$ may be evaluated by
expanding the ground state
$|0 \rangle$ in Eq.~(\ref{greenf}) to ${\cal O}((t/U)^{L-1})$.
The resulting expression for the leading
contribution to $t[G_{\uparrow \uparrow}(L-1)+G_{\downarrow
\downarrow}(L-1)]$ is identical to that for
$2D_{\rm c}(L)/L$ in Eq.~(\ref{bw}),
except that $|0_{\infty} \rangle$ now refers to the infinite system, not the
periodic system of length $L$ \cite{bandw}.  One can fix the phase
factors such that $G_{\uparrow \uparrow}=G_{\downarrow \downarrow}$.
Then the only difference between $tG_{\sigma \sigma}(L-1)$ and
$D_{\rm c}(L)/L$ in the large-$U$
limit stems from the difference between the spin correlations in the infinite
Heisenberg model and those in the periodic one of size $L$, which can at most
cause $tG_{\sigma \sigma}(L-1)$
to differ from $D_{\rm c}(L)/L$ by an algebraic factor in $L$.
The {\em correlation lengths} defined by the two functions must be the
same in the large-$U$ limit.

We have thus established
Eq.~(\ref{corr}) in both the weak- and strong-coupling limits.
We now give a physical argument which suggests that $G$ has this asymptotic
form for arbitrary $U$.

\subsection{Singlet Pairing for $U<0$}
\label{attractive}

The modulus
$|G_{\sigma \sigma}(|x|)|$
of the
equal-time single-particle Green's function
is invariant
under $U\rightarrow -U$ at half filling, as may be seen by applying the
transformation (\ref{e-h}) to Eq.~(\ref{greenf}).
Let us therefore consider Eq.~(\ref{corr}) for the case $U<0$.
The $U<0$ Hubbard model is a superconductor (or more
precisely, in one dimension, has divergent superconducting
fluctuations \cite{korepin}).
Superconductors are characterized by a correlation length $\xi$ which
one may think of as the size of a Cooper pair, and $G$
decays exponentially with characteristic length $\xi$.
The superconducting
correlation length can also be determined
via the periodicity of the ground state energy of a superconducting ring
with respect to the magnetic
flux
enclosed.
Gauge invariance implies that $E_0$ must be
periodic in $\Phi_{\rm c}$ with period $2\pi$ (one flux quantum)
\cite{byersyang};
however, in a system with superconducting
correlations, the period will be $\pi$ (half
a flux quantum) when the system is large compared to the
correlation length
\cite{byersyang}.
In this subsection, we show that the deviations
from a periodicity of
half a flux quantum in the ground state energy of the $U<0$ 1D Hubbard
model
scale as $\exp[-L/\xi(|U|)]$ at $n=1$.  We therefore
identify $\xi(|U|)$ as
the superconducting correlation length;
this makes it plausible that $G$ decays exponentially with this length
for arbitrary $U$,
as asserted in Eq.~(\ref{corr}).
The analogous electron-hole correlations
in the ground state of
the Mott insulator were discussed by Krishnamurthy {\em et al}.\
\cite{jayaprakash} in a slightly different context.

The Bethe ansatz description of the
ground state of the attractive 1D Hubbard model
involves complex pseudomometa \cite{sutherland}, and is unwieldy
for a finite system with periodic boundary conditions.
However, we can use the identity
(\ref{negu}) to obtain the ground state energy of the half-filled
attractive 1D Hubbard model as a function of $\Phi_{\rm c}$ from the ground
state energy of the half-filled repulsive 1D Hubbard model as a function
of $\Phi_{\rm s}=(\Phi_{\uparrow}-\Phi_{\downarrow})/2$.
The latter may be obtained from a solution of
Eqs.~(\ref{keqn}) and (\ref{leqn}) with the $k$'s and $\Lambda$'s
strictly real (alternatively, one can think of this as a Bethe ansatz
for $U<0$ in a representation where the down-spin electrons are treated as
holes).
In 1D, the variation of the ground state
energy with respect to flux for $U<0$ is
$E_0 (\Phi_{\rm c})\simeq E_0 (0) +
D_{\rm c}(-|U|)\Phi_{\rm c}^2/L$ when $L$ is large.  However, there are level
crossings with states whose energy parabolas have local minima at
$\Phi_{\rm c}= \pm \pi, \pm 2\pi$, etc.  Each parabola corresponds to a
particular choice of the Bethe ansatz quantum numbers $\{I_n\}$ and
$\{J_{\alpha}\}$ (in the repulsive representation).
The quantum numbers $\{I_{n}^0\}$ and
$\{J_{\alpha}^0\}$ of the ground state at $\Phi_{\rm c}=0$
are given in Eqs.~(\ref{ieqn}) and (\ref{jeqn}) in
Appendix~\ref{calcxi}.
The ground state at $\Phi_{\rm c}=2\pi$ has quantum numbers satisfying
$I_n=I_{n}^0+1$, $J_{\alpha}=J_{\alpha}^0-2$,
and is equivalent to the above state, but shifted by one flux quantum.
The equivalence of these two states follows from gauge invariance
\cite{byersyang}.
There is a level crossing with a third, inequivalent state whose energy
minimum lies
at $\Phi_{\rm c}=\pi$, and whose quantum numbers satisfy $I_n=I_{n}^0$,
$J_{\alpha}=J_{\alpha}^0-1$.  The energies of these three states as a
function of $\Phi_{\rm c}$ are
plotted as solid curves in
Fig.~\ref{fluxquantum} for $U=-4$ and $N=L=6$ and 28 (note
that $\xi(4)\approx 4.06$).  The ground state energy for the repulsive
case (shifted by $-N|U|/2$) is plotted as a dashed line for comparison.
Equations~(\ref{keqn}), (\ref{leqn}), and
(\ref{eperiodic}) imply that in the large-$L$ limit the local energy
minimum at $\Phi_{\rm c}=\pi$ differs from that at $\Phi_{\rm c}=0$ by
\FL
\begin{equation}
E_0(\pi)-E_0(0) = \frac{4D_{\rm c}(|U|)}{L} \sim
L^{-1/2}\exp[-L/\xi(|U|)].
\label{fluxeq}
\end{equation}
It is evident from Fig.~\ref{fluxquantum} that as $L$ increases and the
local energy minima at $\Phi_{\rm c}=0$ and $\Phi_{\rm c}=\pi$
become degenerate, the
ground state energy becomes periodic with period $\pi$ (half a
flux quantum).

\section{SCALING BEHAVIOR OF $D_{\rm c}$ NEAR THE\\
METAL-INSULATOR CRITICAL POINT}
\label{criticalscaling}

We have shown that the charge response of the half-filled ($n=1$) $U>0$
1D Hubbard model may be characterized by a length $\xi$ which diverges as
$U \rightarrow 0$.  In this section, we show that the point $U=0$, $n=1$
is a conventional quantum critical point in the sense that the singular
behavior of $D_{\rm c}(n,L,U)$ in the vicinity of this point is given by
a (presumably universal \cite{kimweichman}) scaling function
$Y(\xi|1-n|,\xi/L)$, which we calculate numerically and, in various
limits, analytically.  Our results confirm the applicability of the
hyperscaling ansatz to this system.
We also discuss the total optical spectral weight $\pi N_{{\rm tot}}$.

The hyperscaling ansatz for a $T=0$ phase transition is that the
singular part of the ground state energy per correlation volume
scales as the inverse of the correlation
time \cite{kimweichman}.  Since the correlation time scales as $\xi^z$
\cite{hertz}, it follows that
\begin{equation}
E_{0}^{\rm sing}/L^d \sim \xi^{-(d+z)},
\label{hyperscalener}
\end{equation}
where $d$ is the spatial dimension
and $z$ is the dynamic critical exponent.
In our problem $d=1$, and the
Lorentz invariance of the critical point implies $z=1$.
 From Eq.~(\ref{stiff}), we see that $D_{\rm c}$ involves two derivatives
of $E_0/L$ with respect to $\Phi_{\rm c}/L$; the singular part of
$D_{\rm c}$ should
therefore have scaling dimension $2-d-z=0$.  We thus postulate that
as $U\rightarrow 0$, $n\rightarrow 1$, and
$L\rightarrow \infty$,
the singular part of $D_{\rm c}$
is given by
\begin{equation}
D_{\rm c}^{\rm sing}(n,L,U)= Y_{\pm}(\xi \delta,\xi/L),
\label{hyperscale2}
\end{equation}
where $\delta=|1-n|$ and $Y_{\pm}$ are dimensionless scaling functions,
universal up to a metric factor
fixing the units of length and energy.
The subscripts $\pm$ on $Y$ refer to $U>0$ and $U<0$.
The charge response at $U=0^-$ is that of noninteracting electrons with
nearest neighbor hopping, so $Y_-(0,0)= 2/\pi$.
One
consequence of Eq.~(\ref{hyperscale2}), therefore,
is that the charge stiffness at
$n=1$ and $L=\infty$ has
a universal jump of $2/\pi$ as $U$ goes from $0^+$ to $0^-$.

Eqs.~(\ref{stiffans}) and (\ref{limitcoeff}) imply that
\begin{equation}
Y_+(0,y)=  (2/\pi y)^{1/2} \exp (-1/y); \;\;\;\;\; y \rightarrow 0.
\end{equation}
To verify the scaling law (\ref{hyperscale2}) for the general case, we have
calculated $D_{\rm c}(n,L,U>0)$
by solving the Bethe ansatz equations (\ref{keqn})
and (\ref{leqn}) numerically for systems with $N\leq 200$ and $L\geq N$.
We have considered only systems with an even number
of electrons $N$.  In order to obtain for $D_{\rm c}$ a smooth function of
$n$ and $L$, we impose antiperiodic boundary conditions
({\em i.e.}, $\Phi_{\uparrow}=\Phi_{\downarrow}=\pi$) when $N\bmod 4=0$,
so that the ground state is always a singlet
\cite{ogatashiba,liebmattis}.
Because $\delta \equiv |1-N/L|= 0$, $1/L$, $2/L, \ldots$, the scaling
function $Y(x,y)$ is only defined on a countable set of lines in the
$x,y$ plane, namely on $x=my$; $m=0$, 1, $2,\ldots, \infty$.
In Fig.~\ref{fig.hyperscaling}, we plot $\pi D_{\rm c}$ versus $\xi/L$ along
the lines $\delta=0$, $1/L$, $2/L$, and $4/L$ for systems with $N=60$,
80, and 100 electrons.  In each case,
the results for $N=60$, 80, and 100 fall onto
a smooth curve over the entire range $0.3 \leq \xi/L \leq 10000$,
verifying Eq.~(\ref{hyperscale2}).
(The limit $L\gg \xi$ is examined more fully in
Sec.~\ref{finite}.)
For comparison, Fig.~\ref{fig.hyperscaling}(b) is replotted with $\xi$
as the abscissa
in Fig.~\ref{fig.hyperscaling}(e).
It is evident from Fig.~\ref{fig.hyperscaling}
that as $\xi/L\rightarrow \infty$, $D_{\rm c}\rightarrow
2/\pi$ (the value of the universal jump at the critical point),
which we may express as
\begin{equation}
\lim_{y\rightarrow\infty} Y_+(x,y)=\frac{2}{\pi}.
\label{universalpt}
\end{equation}

The charge stiffness in the limit
$L \rightarrow \infty$ ($y\rightarrow 0$)
has been calculated previously
\cite{schulzcond,kaw} from a numerical solution of the Bethe ansatz
integral equations.
Here we merely point out that because $Y$ is universal,
one can extract
$Y_+(x,0)$ from Haldane's analytical
result \cite{haldane1} for the sine-Gordon model
at $\beta^2=8\pi^-$.  The mapping between the two models
is uniquely
determined by equating the energy gap $\Delta$ to the soliton rest
energy $m_{\rm s} c^2$, equating
the correlation length $\xi$ to the quantum soliton length, and equating
the doping $\delta=|1-n|$ to the mean number of solitons per unit
length, as discussed in Sec.~\ref{greenxi}.
The result for the scaling function is \cite{haldane1}
\FL
\begin{equation}
Y_+(x,0) = x
\left(1-\frac{1}{2}(\pi x)^2 + \frac{8\ln 2}{3\pi^2}(\pi x)^3
+ \cdots \right);\;\;\;\;\;
x \rightarrow 0,
\label{stiffhaldane1}
\end{equation}
\begin{equation}
Y_+ (x,0) = \frac{2}{\pi} \left(1 - \frac{1}{2}[\ln(\pi x)]^{-1} +
\cdots \right); \;\;\;\;\;
x \rightarrow \infty.
\label{stiffhaldane2}
\end{equation}
Near the critical point, therefore, the charge stiffness
grows rapidly as one dopes the Mott insulator,
saturating near its maximum
value $2/\pi$ when $x=\xi \delta \sim 1$.
In Sec.~\ref{sec.holonband}, we compute the small-$x$ behavior of
the scaling function directly from the Bethe ansatz equations, verifying
Eq.~(\ref{stiffhaldane1}).

The total optical spectral weight $\pi N_{{\rm tot}}$
involves information
at high frequencies, but is related via the {\em f}-sum rule
(\ref{total}) to the expectation value of the
kinetic energy in the ground state; thus $N_{\rm tot}$ may also be
expected to exhibit universal behavior near the critical point.
In the critical regime $\xi$, $L$, $\delta^{-1} \gg 1$, we find
that \cite{refschulz}
\begin{equation}
N_{{\rm tot}}\simeq \frac{2}{\pi},
\label{totalscale}
\end{equation}
approximately independent of $\delta$, $L$, and $\xi$.
The approximate $\delta$-independence of $N_{\rm tot}$ near the critical
point is in agreement with the intuitive picture that when $\xi \gg 1$
the kinetic energy of the system should be little affected by the
transition to an insulating state.
At $\delta=0$, there is an energy gap $2\Delta$
for charge excitations, and all the optical
spectral weight is at frequencies $\omega \geq 2\Delta$ when $L=\infty$.
We refer to the states at $\omega \geq 2\Delta$ as being
in the ``upper Hubbard band'' (UHB) (see Sec.~\ref{sec.holonband}).
Although for $\delta > 0$ there is no
true gap in $\sigma(\omega) \equiv
\mbox{Re}\,[\tilde{\sigma}(\omega)]$, the structure in the
UHB region $\omega \geq 2\Delta$ must persist
for sufficiently small $\delta$ by continuity.
However, the rapid growth of $D_{\rm c}$ with $\delta$ and
the approximate $\delta$-independence of $N_{\rm tot}$ imply that
near the critical point the primary effect of doping
on $\sigma(\omega)$ is to
transfer spectral weight from the UHB region into the
Drude peak at $\omega=0$, so that the UHB structure in $\sigma(\omega)$
is essentially destroyed for $\delta > \xi^{-1}$.
A similar shift of spectral weight from the UHB
region into the Drude peak with decreasing system size is implied by
Fig.~\ref{fig.hyperscaling}, the UHB
structure being essentially destroyed for $L < \xi$.
In the language of the continuum field theory \cite{emery},
the approximate equality $N_{{\rm tot}} \simeq D_{\rm c}\simeq 2/\pi$,
which holds in the weak-coupling limit both for $\xi \delta \gg 1$
and for $\xi \gg L$, arises because the effects
of umklapp scattering are largely smoothed out in those limits, so that
$\tilde{\sigma}(\omega)$ approaches the result expected in the absence
of umklapp scattering \cite{giamarchi}, with $N_{{\rm tot}} =
D_{\rm c} = (2/\pi) \sin (n\pi/2)$.

\section{Finite-Size Corrections to the Optical Spectral Weights}
\label{finite}

In this section, we calculate
the finite-size corrections (for $L>\xi$)
to the charge stiffness
$D_{\rm c}$ and the total optical spectral weight $\pi N_{{\rm tot}}$ in the
metallic
phase of the $U>0$ 1D
Hubbard model.
Since the scaling function for the charge stiffness is expected to be
universal \cite{kimweichman},
the finite-size corrections we
obtain should be relevant for other, nonsoluble models as well.

We find that the finite-size corrections to both
$D_{\rm c}$ and $N_{{\rm tot}}$
are positive and depend smoothly on $L$, provided
the boundary conditions are chosen so that the
ground state with an even number
of electrons is a singlet:  periodic boundary conditions when $N \bmod 4=2$,
anti-periodic when $N \bmod 4 = 0$ \cite{boundcond}.
We employ these boundary conditions
throughout this section.
In Fig.~\ref{finitesize}, $D_{\rm c}$ and $N_{{\rm tot}}$
are displayed as a function
of $L$ for $\delta=0.2$ and
$U=6$ ($\xi \approx 2.04$).
Note that the finite-size corrections to
$D_{\rm c}$ are larger than those to
$N_{{\rm tot}}$; this is a generic feature of the
finite-size corrections, and indicates that the finite-size enhancement of
$D_{\rm c}$ is in part due to a transfer of
spectral weight from higher frequencies,
as discussed above.
We find that the leading finite-size corrections to both $D_{\rm c}$ and
$N_{{\rm tot}}$ are ${\cal O}(L^{-2})$ in the metallic phase \cite{Lminus2},
in accordance with the system-size dependence of the ground state
energy \cite{woynarovich2}.
The solid curves in Fig.~\ref{finitesize}
are extrapolations of the $L^{-2}$ behavior to small $L$, indicating that
the asymptotic form of the finite-size corrections gives a reasonable
approximation even for fairly small systems, provided $L>\xi$.

The coefficients $A(\delta,U)$ and $B(\delta,U)$ of the leading
finite-size corrections in the metallic phase, defined by
$D_{\rm c}(L) = D_{\rm c}(\infty) [1+ A(\delta,U)/L^{2} + \cdots]$ and
$N_{{\rm tot}}(L) = N_{{\rm tot}}(\infty)
[1 + B(\delta,U)/L^{2} + \cdots]$
are plotted
versus $U$ for several values of $\delta$ in Fig.~\ref{largesize}
(note the different vertical scales).
Both for
$U\rightarrow 0$ and far from the critical point,
$A(\delta,U)$, $B(\delta,U) \rightarrow
\pi^{2}/6$, the value expected for noninteracting fermions (with or
without spin) with nearest-neighbor hopping.
While the relative finite-size corrections to the total optical
spectral weight are always of order
$\pi^{2}/6L^{2}$, the relative finite-size corrections to $D_{\rm c}$
can be much larger, particularly near the critical point.
The curves in Fig.~\ref{largesize}(a) display increasingly sharp maxima
as $\delta$ is decreased, which occur at values of $U$ such that $\xi
(U) \delta \approx 0.4$, and at which point $A(\delta,U) \approx 0.8
\xi(U)^2$.

In the critical regime, we expect the leading finite-size corrections
to $D_{\rm c}$ to scale as
\FL
\begin{equation}
D_{\rm c}(L) = D_{\rm c}(\infty)
[1 + \pi^{2}/6L^{2} + f_+(\xi \delta) (\xi/L)^{2}
+ \cdots],
\label{sizescale}
\end{equation}
where $f_+(x)=d^2Y_+(x,y)/dy^2|_{y=0}/2Y_+(x,0)$, and we have included the
regular term $\pi^2/6L^2$.
The small-$x$ behavior of $f_+(x)$ can be obtained by
noting that to ${\cal O}(\xi\delta)^3$, the charge stiffness in the
weak-coupling limit is equivalent to that of noninteracting relativistic
spinless fermions \cite{haldane1} [{\em c.f.}
Eq.~(\ref{stiffhaldane1})], whence
\begin{equation}
f_+(x) = \frac{\pi^{2}}{2} -\pi^{4} x^{2} + \cdots;\;\;\;\;\;x\rightarrow
0.
\label{f0}
\end{equation}
In Fig.~\ref{scaling}, the function
$f_+\equiv [A(\delta,U)-\pi^2/6]
/\xi(U)^2$
is plotted versus
$\xi(U)\delta$;
the solid curve was obtained by fixing $\delta$ and varying
$U$, while the triangles were obtained by fixing $U$ and varying $\delta$.
Note that the numerical data connect smoothly to Eq.~(\ref{f0}) (dotted curve)
as $\xi\delta \rightarrow 0$.  Fig.~\ref{scaling} provides further
proof of the scaling law (\ref{hyperscale2}) in the regime $L\gg
\xi$.

Though the finite-size corrections to $D_{\rm c}$ near the critical point are
much larger than expected in a noninteracting system, they should not be a
great obstacle to obtaining information about optical properties of 1D
models from numerical calculations on small lattices, except at $\delta =0$,
due to the impossibility of simultaneously having small $L$ and small but
finite $\delta$ ($\min \delta =1/L$).  Roughly speaking, the maximum value
of the $L^{-2}$ correction in the critical regime is
\begin{equation}
\frac{D_{\rm c}(L)-D_{\rm c}(\infty)}{D_{\rm c}(\infty)} \approx
\frac{0.13}{(\delta L)^{2}},
\label{error}
\end{equation}
though of course we've seen in Fig.~\ref{finitesize} that the additional term
$\pi^2/6L^2$ can lead to slightly larger corrections
($\sim 20\%$) for small systems.

A further comment on the choice of boundary conditions is appropriate at this
point.  One practice which has been employed to reduce finite-size effects
is to average over periodic and antiperiodic boundary conditions.  While this
may result in cancellations for some values of $n$ and $U$ due to
the fact that the finite-size
corrections sometimes have different signs, it is
ineffective near the critical point, where the
finite-size corrections are largest, since the corrections are always positive
and essentially independent of boundary conditions in that regime
\cite{exception}.  Furthermore,
depending on $U$ and $n$, any of the three possibilities---either
periodic or antiperiodic boundary conditions or an average over
the two---can result in the smallest finite-size corrections.
One is better off choosing boundary conditions such that
the ground state with an even number of electrons is a singlet; then the
finite-size corrections to $D_{\rm c}$ and
$N_{{\rm tot}}$ are always positive,
with magnitudes as given above.

\section{CHARGE EXCITATIONS AND THERMOPOWER NEAR\\ THE METAL-INSULATOR
TRANSITION}
\label{sec.holonband}

The mapping of the charge sector of the weak-coupling 1D Hubbard model
near $n=1$
onto the sine-Gordon model suggests that the charge excitations near
half filling
can be described
in terms of the states of $L-N$ soliton-like
charge carriers which behave like weakly
interacting spinless fermions in the low-density limit $L-N \ll L$.
In this section, we use a reformulation of the Bethe ansatz equations
due to Woynarovich \cite{woy4} to construct an explicit mapping for arbitrary
$U$ and $\xi \delta \ll 1$ of the
low-lying charge degrees of freedom of the 1D repulsive-interaction
Hubbard model onto spinless fermions whose
mutual interactions vanish as $\xi \delta \rightarrow 0$,
and which obey a quadratic energy-momentum dispersion relation with a
$U$-dependent effective mass.  We use this mapping to obtain
asymptotically exact expressions for the charge stiffness and
low temperature thermopower near the metal-insulator
transition.  Our calculation of the thermopower extends
and makes more rigorous previous work of Schulz \cite{schulztherm} which
was based on a weak-coupling approximation,
and implies that the transport is hole-like for $0<1-n\ll\xi^{-1}$.

The energy eigenstates of
a $U>0$ Hubbard chain of length $L$ containing
$L-H$ electrons can be obtained by solving Eqs.~(\ref{keqn}) and
(\ref{leqn}), with an appropriate choice of charge quantum numbers
$\{I_n\}$ and spin quantum numbers $\{J_{\alpha}\}$.  The charge
quantum numbers of the ground state are $L-H$ consecutive integers (or
half-integers), centered about the origin, and the spin quantum numbers
of the ground state are $(L-H)/2$
consecutive integers (or half-integers), also centered about the origin
(as discussed in
Sec.~\ref{form}).
We wish to consider the low-lying charge
excitations of the system, so we restrict our attention
to those states whose spin quantum numbers $\{J_{\alpha}\}$ are those of
the ground state, 
and in which all of the pseudomomenta are real (states
with complex pseudomomenta are separated from these by an energy gap
\cite{woy4}).
A hole $I_h$ in the distribution of the charge quantum numbers ({\em
i.e.}, $I_{n+1}-I_n=2$, $I_h \equiv I_n+1$) corresponds to a charge
excitation which, following Anderson \cite{phil3}, we refer to as a {\em
holon}.
We wish to consider the case $H\ll L$;  since
the $I_n$ are only defined modulo($L$), we can characterize the
low-lying charge excitations of the system by the $H$ holes $\{I_h;
h=1,\ldots,H\}$ in the distribution of the charge quantum numbers,
rather than working with the larger set $\{I_n\}$.  Formally, we shall
consider these holes in the distribution of the charge quantum numbers
to exist even in the ground state of the system,
where they will be consecutive integers (or half-integers) centered
about $L/2$ (for convenience, we restrict $I_h$
to the region $1 \leq I_h \leq L$).
We denote the corresponding holes in the distribution of the
pseudomomenta by the set $\{k_h\}$.
Woynarovich \cite{woy4} has derived a reduced form of the Bethe ansatz
equations, valid in the large-$L$ limit,
which deals only with the parameters $\{I_h\}$ and $\{k_h\}$ of these
excitations,
rather than the parameters of all the electrons.
We extend the results of Ref.~\cite{woy4}
to include a magnetic flux
$\Phi_{\uparrow}=\Phi_{\downarrow}=\Phi_{\rm c}$ through
the ring, and point out that the reduced set of Bethe ansatz equations
derived in Ref.~\cite{woy4} can be interpreted as an asymptotic
Bethe ansatz for {\em holons}, which consequently is expected to become
exact in the limit $L \rightarrow \infty$ for arbitrary $H$.
We use this asymptotic Bethe ansatz to
investigate in detail the limit $H$, $L \rightarrow \infty$,
with $\delta=H/L$ small but finite.

In the large-$L$ limit,
the energy of such a state is given in terms of the set
$\{k_h\}$ by \cite{woy4}
\begin{equation}
E(L-H)=E_0(L) -\sum_{h=1}^{H} \varepsilon_{\rm c}(k_h), 
\label{enerholes}
\end{equation}
where $E_0(L)$ is the ground state energy at half filling
\cite{lwu},
\begin{equation}
E_0(L) = -4 L \int_{0}^{\infty} \frac{J_0(\omega) J_1(\omega)
\,d\omega}{\omega [1 + \exp(\omega U/2)]},
\label{liebwuener}
\end{equation}
and \cite{woy4}
\FL
\begin{equation}
\varepsilon_{\rm c}(k) = -2\cos k -4 \int_{0}^{\infty} \frac{J_1(\omega) \cos
(\omega \sin k)\, d\omega}{\omega [1+\exp(\omega U/2)]}.
\label{chargener}
\end{equation}
The momentum of the state, defined modulo($2\pi$), is \cite{woy4}
\begin{equation}
P = \Phi_{\rm c} - \sum_{h=1}^{H} p_{\rm c}(k_h) + \pi(L-N/2+1),
\label{pholes}
\end{equation}
where
\begin{equation}
p_{\rm c}(k)
= k + 2 \int_{0}^{\infty} \frac{J_{0}(\omega)
\sin (\omega \sin k)\,d \omega}{\omega[1 + \exp(\omega U/2)]}.
\label{pofk}
\end{equation}
Following Anderson \cite{phil3},
we interpret $-\varepsilon_{\rm c}(k_h)$ and
$-p_{\rm c}(k_h)$ as the energy and momentum of a {\em holon}.
The holons can not be regarded as noninteracting quasi-particles,
however, since the
$k_h$ are not free parameters, but are related to the set $\{I_h\}$
by the equations \cite{woy4}
\FL
\begin{equation}
L p_{\rm c}(k_h) = 2\pi I_h + \Phi_{\rm c} +
\sum_{h'=1}^{H} \Theta (k_h,k_{h'}),
\label{kheqn}
\end{equation}
where
\begin{equation}
\Theta (k,k')=
2 \int_{0}^{\infty} \frac{\sin[\omega(\sin k - \sin k')] \, d\omega}{
\omega[1+\exp(\omega U/2)]}.
\label{gamma}
\end{equation}

We point out that
the set of equations (\ref{kheqn}) is formally equivalent to an
{\em asymptotic 
Bethe ansatz} for the holons,
$\Theta(k,k')$ being the effective holon-holon scattering phase shift
\cite{phaseshift}.  An asymptotic Bethe ansatz \cite{sutherland}
can be used to obtain
the asymptotic form of the many-body wavefunction of a 1D system with
non-diffractive scattering in the limit where the particles are
widely separated---even if the exact wavefunction of
the system is not of Bethe ansatz form.
Moreover, the asymptotic form of
the wavefunction is sufficient to determine the energy eigenvalues of
the system
{\em exactly} in the limit $L\rightarrow \infty$, even at finite
particle density \cite{sutherland}.
Consequently, we expect Eqs.~(\ref{enerholes}),
(\ref{pholes}), and (\ref{kheqn}) to become exact in the
limit $L\rightarrow \infty$ for arbitrary $H$.
We emphasize that this conclusion follows {\em a posteriori} from the
form of the equations, and is by no means obvious from the method
\cite{woy4} of their derivation.
As a test of this conjecture, we use
Eqs.~(\ref{enerholes}) and (\ref{kheqn}) to calculate the
energy of the empty lattice, which in this formalism is treated as
the ground state of a system of $L$ holons.  In the limit
$H=L\rightarrow \infty$,
Eq.~(\ref{kheqn}) implies that the holon
pseudomomenta $\{k_h\}$ are equally spaced with density $L/2\pi$, so
that the sum over $k_h$ in Eq.~(\ref{enerholes}) can be
replaced by $L\int_{-\pi}^{\pi}\,dk/2\pi$.  The resulting integral is
equal to $E_0(L)$, yielding $E=0$, as expected for an empty lattice.

Equations~(\ref{chargener}) and (\ref{pofk}) implicitly define an
energy band $\varepsilon_{\rm c}(k(p))$ for charge excitations, $k(p)$ being
the inverse of the function $p_{\rm c}(k)$ defined in Eq.~(\ref{pofk}).  The
holons are just holes in this energy band, which is full at $N=L$.
Since $d\varepsilon_{\rm c}(k(p))/dp|_{p=\pi}=
d^3\varepsilon_{\rm c}(k(p))/dp^3|_{p=\pi}=0$,
the holon energy near the zone boundary $p=k=\pi$ can be written
\begin{equation}
\varepsilon_{\rm c}(k(p)) = \mu_- - \frac{(p-\pi)^2}{2|m^{\ast}|}
+ \frac{\lambda (p-\pi)^4}{4!} + \cdots,
\label{holonband}
\end{equation}
where $\mu_- =\varepsilon_{\rm c}(\pi)$ is the chemical potential in the limit
$n \rightarrow 1^-$ \cite{lwu},
\begin{equation}
\frac{1}{m^{\ast}} \equiv \left.\frac{d^2 \varepsilon_{\rm c}(k(p))}{dp^2}
\right|_{p=\pi} = \frac{\varepsilon_{\rm c}''(\pi)}{[p_{\rm c}'(\pi)]^2} < 0,
\label{mdef}
\end{equation}
and
\begin{equation}
\lambda \equiv \left.\frac{d^4 \varepsilon_{\rm c}(k(p))}{d
p^4}\right|_{p=\pi} = \frac{\varepsilon_{\rm c}^{(4)}(\pi)}{[p_{\rm
c}'(\pi)]^4} -  \frac{4 \varepsilon_{\rm c}''(\pi) p_{\rm c}^{(3)}(\pi)}{
[p_{\rm c}'(\pi)]^5}.
\label{fourth}
\end{equation}
Eq.~(\ref{mdef}) defines
the effective mass $m^{\ast}$ of the holons near the zone boundary
\cite{kawokiji,carmelo,thesis},
which is negative, as is appropriate for a hole-like charge
carrier.  Explicit expressions for $p_{\rm c}'(\pi)$, $\varepsilon_{\rm
c}''(\pi)$, $p_{\rm c}^{(3)}(\pi)$, and $\varepsilon_{\rm c}^{(4)}(\pi)$
are given in Eqs.~(\ref{pfirst})--(\ref{efourth}) in Appendix \ref{appB}.

Eq.~(\ref{kheqn}) implies that the holon momenta differ from those of
noninteracting spinless fermions by a term which vanishes as $\delta
\rightarrow 0$; we write $p_h \equiv
p_{\rm c}(k_h) = 2\pi I_h/L + \Phi_{\rm c}/L +
\delta p_h$, where $\delta p_h = L^{-1}\sum_{h'=1}^{H}
\Theta(k_h,k_{h'})$.  Eq.~(\ref{gamma}) implies that
\begin{equation}
\lim_{k_h,\,k_{h'} \rightarrow \pi}
\Theta(k_h, k_{h'}) = -\frac{4 \ln 2}{U p_{\rm c}'(\pi)} \left[
p_{\rm c}(k_h) - p_{\rm c}(k_{h'}) \right].
\label{thetasim}
\end{equation}
For small $\delta =H/L$ and low temperatures [$k_B T \ll E_F \simeq (\pi
\delta)^2 / 2 |m^{\ast}|$], we can use this
expression for $\Theta$ to calculate $\delta p_h$:
since the thermal average of $2\pi I_{h'}/L$ is $\pi$, the thermal
average of $\delta p_h$ (for fixed $I_h$) is
\begin{equation}
\langle\langle \delta p_h \rangle\rangle
= - \frac{4 \ln 2 \, \delta}{U p_{\rm c}'(\pi)}
\left(\frac{2 \pi I_h}{L} - \pi \right) + {\cal O}(\delta^4).
\label{deltaph}
\end{equation}
Inserting this result into Eq.~(\ref{holonband}), one finds
that the shift of the holon momentum $-\delta p_h$
caused by holon-holon scattering leads to a fractional shift in the
holon energy,
measured relative to the chemical potential at half
filling, given by
\begin{equation}
\frac{-\delta \varepsilon_{\rm c}}{\mu_- - \varepsilon_{\rm c}}
\simeq -\frac{8 \ln 2 \, \delta}{U p_{\rm c}'(\pi)} = \left\{
\begin{array}{ll}
-4 \ln 2 \, \xi \delta / \pi \;\;\;\; & U \rightarrow 0 \\
-8 \ln 2 \, \delta / U & U \rightarrow \infty.
\end{array}\right.
\label{holoninteractions}
\end{equation}
The effective holon-holon interaction, which should be considered as a
correction to the hard-core repulsion of free spinless fermions, is thus
attractive, and is negligible for $\xi
\delta \ll 1$ (or $U \gg t$).  While the explicit form of
Eq.~(\ref{holoninteractions}) is only valid for $\xi \delta \ll 1$ and
small excitation energies, one can show quite generally that $|\delta
p_h| < 8 \ln 2 \, \delta / U$, so that holon-holon interactions are
also negligible at high temperatures when $\delta / U \ll 1$.

The ground state of the system is obtained by choosing the set $\{I_h\}$ to
be consecutive integers (or half-integers) centered about $L/2$, {\em
i.e.}, by placing the holons near the energy minimum at
$p= \pi$ (with our convention for $I_h$, $\,p_h$ is restricted
to the interval $0 \leq p_h \leq 2 \pi$).
Since $\delta p_h/p_h \sim {\cal O}(\xi \delta)$ as
$\delta \rightarrow 0$,
the holon momenta are approximately equally spaced in the ground state
when $\xi\delta \ll 1$.
We can thus use Eqs.~(\ref{enerholes}), (\ref{kheqn}), and (\ref{holonband})
to obtain an analytic expression for the
charge stiffness near half filling \cite{kaw2},
\begin{equation}
D_{\rm c} = \frac{\delta}{2|m^{\ast}|} - \frac{\pi^2 \lambda \,
\delta^3}{12} +
\frac{2 \pi^2 \ln 2\, \delta^4}{3\, U p_{\rm c}'(\pi)}
\left(\lambda + \frac{p_{\rm c}'(\pi) + p_{\rm c}^{(3)}(\pi)}{|m^{\ast}|
[p_{\rm c}'(\pi)]^3} \right) + {\cal O}(\delta^5).
\label{stiffholon}
\end{equation}
To ${\cal O}(\delta^3)$, the
charge stiffness is equivalent to that of noninteracting spinless
fermions in the energy band (\ref{holonband}),
reflecting the approximate Galilean invariance which holds at low holon
densities.
The term proportional to $\delta^4$ comes from holon-holon interactions.
The term linear in $\delta$ in Eq.~(\ref{stiffholon})
may also be obtained by a more conventional technique, which we
describe in Appendix \ref{appB}.

It is instructive to consider the
weak- and strong-coupling limits of Eq.~(\ref{stiffholon}).
Making an asymptotic expansion of the $U$-dependent coefficients
in Eq.~(\ref{stiffholon}) about $U=0$, and using
Eqs.~(\ref{smallU}) and (\ref{light}), we obtain
\begin{equation}
\lim_{U\rightarrow 0}
D_{\rm c} = \frac{c \xi \delta}{2} \left(1 - \frac{1}{2} (\pi \xi
\delta)^2 + \frac{8 \ln 2}{3 \pi^2}
(\pi \xi \delta)^3 + {\cal O}(\xi \delta)^4 \right),
\label{Dsol}
\end{equation}
which is equivalent to Eq.~(\ref{stiffhaldane1}), apart from the
nonsingular multiplicative factor
$c/2t =1 + U/4\pi t$.
Note that the weak-coupling limit of the holon
effective mass can be written as
$\lim_{U \rightarrow 0} |m^{\ast}| = \Delta/c^{2}$,
which is just the
rest mass of the soliton in the equivalent sine-Gordon model.
In the limit $U\rightarrow \infty$, Eq.~(\ref{stiffholon}) becomes
\begin{equation}
\lim_{U\rightarrow \infty}
D_{\rm c} = \frac{t}{\pi} \left(\pi \delta - \frac{(\pi \delta)^3}{3!} +
{\cal O}(\delta^5) \right),
\end{equation}
where we have made the hopping matrix element $t$ explicit.
This is just the Taylor series for $(t/\pi) \sin \pi \delta$,
the charge stiffness of noninteracting
spinless fermions with dispersion $\varepsilon_{\rm c}(p)=-2t\cos p$,
in accord with the well known behavior of the holons in the limit $U \gg t$
\cite{woy4,ogatashiba}.

The low-lying charge excitation eigenstates of a Hubbard ring with
$L+H'$ electrons follow from the above results and the particle-hole symmetry
present at half filling \cite{woy4}.  The energy and momentum
of such a state are given
by Eqs.~(\ref{enerholes}) and (\ref{pholes})
with $-\varepsilon_{\rm c}(k_h) \rightarrow
U - \varepsilon_{\rm c}(k_h)$,
and the $k_h$ are determined by Eq.~(\ref{kheqn})
with $\Phi_{\rm c} \rightarrow
-\Phi_{\rm c}$.  For $H' \ll L$, and for small
excitation energies, these states can
be described as the states of $H'$ noninteracting spinless fermions
(antiholons) near the bottom of the parabolic energy band,
\begin{equation}
\varepsilon_{\rm c}^{\rm UHB}(p)=
U-\varepsilon_{\rm c}(k(p)) = U-\mu_- + \frac{(p-\pi)^2}{2|m^{\ast}|}
- \frac{\lambda (p -\pi)^4}{4!} + \cdots.
\label{antiholonband}
\end{equation}
The antiholons are thus characterized by a positive effective mass near
$p=\pi$.  Note that the gap between the maximum of the lower Hubbard band
(\ref{holonband}) and the minimum of the upper Hubbard band
(\ref{antiholonband}) is $U-2\mu_- =2\Delta$, the Lieb-Wu charge gap
\cite{lwu,woy4}.  The result (\ref{stiffholon}) also holds for $n=1+\delta$.
We can thus write
\begin{equation}
\frac{\partial D_{\rm c}}{\partial n} = \frac{1}{2 m^{\ast}};
\;\;\;\;\; |1-n| \ll 1,
\label{Drude}
\end{equation}
where it is understood that $m^{\ast} > 0$ ($ < 0$) for $n > 1$ ($ < 1$).
This expression emphasizes that the conductivity is ``hole-like''
when $m^{\ast} <0$, in that the Drude weight {\em decreases}
when an electron is added to the system, and
conversely, that it is ``electron-like'' when $m^{\ast} >0$ \cite{Drudefree}.

The physical significance of the effective mass $m^{\ast}$ is revealed in the
dynamics of a wave packet.
The {\em charge velocity} $v_{\rm c}$ is defined as the
the group velocity of a wave packet composed of low energy charge excitations.
For $\delta \ll 1$, we obtain
\begin{equation}
v_{\rm c} =
\left.\frac{d \varepsilon_{\rm c} (k(p))}{dp}\right|_{p=\pi(1-\delta)}
\simeq \frac{\pi \delta}{|m^{\ast}|}.
\label{chargevel}
\end{equation}
If we apply an electric field ${\bf E}$ to the system, the wave packet
will be accelerated \cite{effcharge}:
\begin{equation}
\frac{d v_{\rm c}}{dt} = \frac{d^2
\varepsilon_{\rm c}}{dp^2}\left(\frac{dp}{dt}\right)=
\frac{1}{m^{\ast}}\left(-e {\bf E}\right).
\label{newtons3rd}
\end{equation}
The holons, having $m^{\ast}<0$, will be accelerated parallel to an applied
electric field, and are thus hole-like, while antiholons are electron-like
\cite{chargedefinitions}.

The transport coefficients which are conventionally used to determine the
sign of the carriers---the thermopower and the Hall coefficient---couple to
electrons, not holons, however, so the above picture is not the whole story.
In 1D, only the thermopower (Seebeck coefficient) $S$ is available, which is
defined in terms of the open circuit electric field $\bf E$ produced by
a temperature gradient $\nabla T$ across the sample:
\begin{equation}
{\bf E}=S \nabla T.
\label{seebeck}
\end{equation}
$S$ is the entropy
carried per unit charge by an electric current \cite{callen},
and is ordinarily negative  for electron conduction and positive for holes.
In general, both the charge and spin entropies will contribute to $S$.
However, the holon density of states diverges ($v_{\rm c}\rightarrow 0$)
as $\delta \rightarrow 0$.  Close to the
metal-insulator transition, therefore, the entropy of the holons
will be much greater than the spin entropy at low temperatures,
and will dominate the thermopower \cite{schulztherm,myprl}.
Furthermore, in this limit the holons
can be treated as noninteracting spinless
fermions in the energy bands (\ref{holonband}) and (\ref{antiholonband}).
So far, we have proven this only for the case where the spin wave
function is in its ground state and there are no charge excitations into the
upper Hubbard band [{\em i.e.}, no excitations from the band (\ref{holonband})
into the band (\ref{antiholonband})].  Such charge excitations
correspond to umklapp processes, and lead to a finite conductivity for
$T>0$ \cite{millisgiamarchi}; however, their
effect on the thermopower is negligible when $k_B T \ll 2\Delta$.
The interaction of the holons with a thermal population of spin excitations
can be described by adding a term $L\,\delta p_{h}^{s}(T)$ to the right
hand side of Eq.~(\ref{kheqn}); in Ref.~\cite{myprl}, we show that
for small $\delta$ and low temperatures $\delta p_{h}^{s}(T) \sim
{\cal O}(T \delta)$,
so the holons are still effectively
noninteracting when $\delta$ and $T$ are small.
A standard formula for the low temperature thermopower \cite{chaikin}
can therefore be used to obtain
\begin{equation}
S=-\frac{k_{B}^{2} T}{3|e|}\frac{m^{\ast}}{\delta^2},
\label{thermopower}
\end{equation}
which is valid for $\xi \delta \ll 1$ and $k_B T \ll E_F \simeq (\pi
\delta)^2 / 2 |m^{\ast}|$ in the absence of impurity scattering.
The low temperature thermopower thus becomes
large and positive as the metal-insulator transition is
approached from $n<1$, and has the opposite sign for $n>1$.
The small-$U$ behavior of Eq.~(\ref{thermopower}) is in agreement with
an earlier result \cite{schulztherm} obtained using a
weak-coupling approximation.
The entropy carried by spin excitations is important
at higher temperatures and/or dopings, and is discussed in
Ref.~\cite{myprl}.

\section{CONCLUSIONS}
\label{conclusions}

We have obtained an exact result for the charge
stiffness of a Hubbard ring with $U>0$ and $n=1$ in the
large-circumference limit, which defines and
yields an analytic expression for the
correlation length $\xi(U)$ in the Mott insulating phase of the 1D
Hubbard model.  We have shown that this correlation length also governs
the exponential decay of the equal-time single-particle Green's function
at $n=1$
in both the weak- and strong-coupling limits, as well as the pairing
correlations in the ground state of the attractive 1D
Hubbard model at $n=1$.  We remark that the strong-coupling expansion
[Eq.~(\ref{bw})] which gave $D_{\rm c}(L)$, $G_{\sigma \sigma}(L)
\sim \exp(-L/\xi)$, with
$\xi^{-1}=\ln(U/t)+\mbox{constant}$,
must hold for the half-filled
Hubbard model in the large-$U$ limit in any spatial dimension.

In the vicinity of the zero temperature critical point $U=0$, $n=1$,
we have shown that the
doping and system-size dependence of the
charge stiffness scale with the correlation length $\xi$.
The scaling function for the charge stiffness is expected to be universal
\cite{kimweichman}, and thus should
be applicable to other 1D metal-insulator
transitions.   In addition, this scaling function appears to describe a
certain class of 1D magnetic phase transitions which do not involve
broken symmetry, and which are characterized by
a transition from a state with algebraic correlations
to a state with a gap.  In particular,
Eqs.~(\ref{negu}) and (\ref{hyperscale2}) imply that the spin stiffness
$D_{\rm s}=(L/2)d^2 E_0/d \Phi_{\rm s}^2|_{\Phi_{\rm s}=0}$
(where $\Phi_{\rm s}=(\Phi_{\uparrow}-\Phi_{\downarrow})/2$)
of the 1D attractive-interaction
Hubbard model with $n=1$ and magnetization
$\sigma =2S^z/L$ has the form
\begin{equation}
D_{\rm s}(S^z,L,-U)=Y_{\pm}(\xi \sigma, \xi/L)
\label{magscaling}
\end{equation}
in the vicinity of the magnetic critical point $U=0$, $\sigma=0$.  The
spin stiffness is related to the magnetic susceptibility $\chi$ by
$D_{\rm s}= (1/2\pi^2)\chi^{-1}$ (see Ref.~\cite{ss}).  There is a
similar mapping between the metal-insulator transition occuring at
$n=1/2$ for 1D lattice
spinless fermions with nearest neighbor repulsion and the magnetic
transition occuring at the isotropic point in the antiferromagnetic
Heisenberg-Ising spin
chain.  Because of the universality of the scaling function for the
metal-insulator transition, this magnetic transition is also expected to be
described by the scaling function $Y$.

One important consequence of scaling is that
the finite-size corrections to $D_{\rm c}$ are enhanced
near the critical point.
While the scaling form (\ref{sizescale})
for the finite-size corrections can only
be expected to hold for 1D systems, it is plausible that the
finite-size corrections to $D_{\rm c}$ may be enhanced
near half filling for smaller values of $U$
in higher dimensional systems as well, suggesting that caution is
required in interpreting numerical calculations of
$\tilde{\sigma}(\omega)$ on small clusters in this regime.
Enhanced finite-size effects of this type
could explain anamolous negative values for the Drude weight
of the 2D Hubbard model on a 4x4 lattice with 14 electrons
when $U=4$ and 8 \cite{xdiag91}.
We do not expect similar enhancements of the finite-size corrections in the
$t-J$ model, since the constraint of no double occupancy fixes $\xi=0$.

In the metallic phase of the model,
the physical significance of the correlation length is that it
defines the characteristic size of the
charge-carrying solitons, or {\em holons}.
We have shown that a reformulation \cite{woy4} of the Bethe ansatz equations
of the 1D Hubbard model in
terms of the parameters of the charge excitations only is formally
equivalent to an asymptotic Bethe ansatz for holons, and have used this
asymptotic Bethe ansatz to show explicitly for arbitrary $U$ that the
holons are equivalent
to spinless fermions
with mutual interactions that vanish in the low-density limit
$\xi \delta \rightarrow
0$.  We have used this mapping to obtain an expression for the low temperature
thermopower near the metal-insulator transition, which implies hole-like
transport for $0<1-n\ll\xi^{-1}$.

It has been argued
in a series of papers by Anderson \cite{phil4} that the physics of doped Mott
insulators is similar in one and two dimensions.
It is therefore interesting to compare the metal-insulator transition in the
1D Hubbard model with that observed in the
cuprate materials in which
high temperature superconductivity occurs, which are widely regarded to be
quasi-two-dimensional doped
Mott insulators \cite{phil}.
The thermopower of the doped cuprates is generically positive
for hole doping \cite{ishii,ando,kaiser} and negative for electron doping
\cite{hagen,xu,sugiyama},
with a magnitude which increases drastically as the nominal concentration
of doped carriers goes to zero, in qualitative agreement with
Eq.~(\ref{thermopower}).  However, linear-$T$ thermopower has not been
clearly identified in metallic samples \cite{impurities}.
For larger dopings, the
thermopower has an unusual temperature dependence
\cite{ishii,ando,kaiser,hagen,xu,sugiyama,ghchen,cohn}, and will be
discussed in more detail elsewhere \cite{myprl}.

The optical conductivity near the metal-insulator transition in the
doped cuprates \cite{scexp1,uchida,scexp3} also shows qualitative
similarities to that near the Mott-Hubbard metal-insulator transition in 1D
\cite{schulzcond,kaw,thesis}.  In particular, a linear growth of
the Drude weight with doping is observed in the cuprates
for small dopings \cite{scexp1,uchida,scexp3},
while the total integrated spectral weight in the optical
conductivity up to $\sim 3\mbox{eV}$, which includes both the low-frequency
response and the charge-transfer band (or upper Hubbard band) is roughly
constant from the insulating antiferromagnet all the way through the
superconducting phase \cite{scexp1,uchida}.
The primary effect of doping in these compounds is thus to transfer spectral
weight from high to low frequencies.
This transfer of spectral
weight is quite rapid, being essentially complete for a doping of $\sim 0.25$
\cite{scexp1,uchida}.
Similarly, near
the critical point of the Mott-Hubbard metal-insulator transition
in 1D, we find that the total optical spectral weight $\pi N_{{\rm tot}}$
is approximately independent of doping,
while the Drude weight $\pi D_{\rm c}$ grows linearly for small dopings,
the transfer of spectral weight from high to low frequencies being
essentially complete when $\xi \delta \sim 1$.
Of course, the fraction of the low-frequency spectral weight which is
collapsed into the Drude peak at $\omega=0$ is much greater in our 1D
model
for several reasons, among which are the absence of an impurity potential,
and the kinematic constraints which limit carrier-carrier scattering in 1D.

\acknowledgments

C.~A.~S. acknowledges support from  AT\&T
Bell Laboratories and NSF grants DMR-91-04873 (at Princeton) and
DMR-91-23577 (at Maryland), and
thanks P.~W.~Anderson, T.~Giamarchi, F.~D.~M.~Haldane,
and B.~S.~Shastry
for valuable discussions.

\appendix{CALCULATION OF $\xi(U)$}
\label{calcxi}

Here we derive the result (\ref{stiffans})
for the asymptotic form of
$D_{\rm c}(L)$ as $L \rightarrow \infty$ at $n=1$
using a technique \cite{woy} previously employed for computing finite-size
corrections to energy eigenvalues
in Bethe ansatz solvable models.

We use Eq.~(\ref{stiff}) to calculate the charge stiffness from the dependence
of the ground state energy on $\Phi_{\rm c}$.
$E_0(\Phi_{\rm c})$ may be obtained by solving the Bethe ansatz equations
(\ref{keqn}) and
(\ref{leqn}),
and using Eq.~(\ref{ener}).
We consider only even $N$ $(=L)$,
so that the ground state is a singlet $(M=N/2)$,
and nondegenerate.  The charge quantum numbers are then:
\begin{equation}
I_{n} = \left\{ -\frac{L}{2}+1-s, -\frac{L}{2}+2-s, \ldots , \frac{L}{2}-s
\right\},
\label{ieqn}
\end{equation}
where $s = \frac{1}{4}(L \bmod 4)$,  and the spin
quantum numbers are:
\begin{equation}
J_{\alpha} = \left\{ -\frac{M-1}{2},
-\frac{M-3}{2}, \ldots, \frac{M-1}{2}\right\}.
\label{jeqn}
\end{equation}
In order to facilitate a solution of Eqs.~(\ref{keqn}) and (\ref{leqn})
in the large-$L$ limit, let us define
the following functions \cite{woy}:
\begin{equation}
p_{L}(k) \equiv k - \frac{1}{L} \sum_{\beta} 2 \tan^{-1}
\frac{\Lambda_{\beta}-\sin k}{U/4},
\label{psubl}
\end{equation}
\FL
\begin{equation}
z_{L}(\Lambda) \equiv \frac{1}{L} \sum_{n} 2 \tan^{-1}
\frac{\Lambda - \sin k_{n}}{U/4} - \frac{1}{L} \sum_{\beta} 2 \tan^{-1}
\frac{\Lambda - \Lambda_{\beta}}{U/2}.
\label{zsubl}
\end{equation}
With these definitions, the Bethe ansatz equations (\ref{keqn}) and
(\ref{leqn}) become,
for $\Phi_{\uparrow}=\Phi_{\downarrow}=\Phi_{\rm c}$,
\begin{equation}
p_{L}(k_{n}) = \frac{2\pi I_{n}}{L} + \frac{\Phi_{\rm c}}{L}, \label{theta}
\end{equation}
\begin{equation}
z_{L}(\Lambda_{\alpha}) = \frac{2\pi J_{\alpha}}{L}.  \label{zed}
\end{equation}
We also introduce the functions \cite{woy}:
\begin{equation}
\rho_{L}(k)
\equiv \frac{1}{2\pi} \frac{d p_{L}(k)}{d k},
\label{rho}
\end{equation}
\begin{equation}
\mu_{L}(\Lambda)
\equiv
\frac{1}{2\pi}\frac{d z_{L}(\Lambda)}{d \Lambda}.
\label{sigma}
\end{equation}
In the limit
$L \rightarrow \infty$ with $N/L\equiv n$ kept finite,
the pseudomomenta $\{k_n\}$ in the ground state
are distributed continuously on the real axis between the
pseudo-Fermi points
$Q_{-}$ and $Q_{+}$ with density
$\rho(k)\equiv {\displaystyle \lim_{L \rightarrow \infty} \rho_L(k)}$, and
the spin rapidities $\{\Lambda_{\alpha}\}$ are distributed continuously
between $-\infty$ and
$\infty$ with density
$\mu(\Lambda) \equiv {\displaystyle \lim_{L\rightarrow \infty}
\mu_L(\Lambda)}$.
Equations~(\ref{keqn}) and
(\ref{leqn}) then
lead to the following coupled integral equations for $\rho(k)$
and $\mu(\Lambda)$ \cite{lwucomment}:
\FL
\begin{eqnarray}
2 \pi \rho(k) & = & 1 + \cos k
\int_{-\infty}^{\infty} d \Lambda \, \frac{8 U  \mu (\Lambda)}{U^{2} +
16(\Lambda - \sin k)^{2}},
\rule[-1cm]{0cm}{1cm}
\label{liebwurho} \\
2 \pi \mu (\Lambda)
& + & \int_{-\infty}^{\infty} d \Lambda' \, \frac{4U  \mu (\Lambda')}{U^{2}
+ 4(\Lambda - \Lambda')^{2}}
 =
\int_{Q_-}^{Q_+} d k \, \frac{8U \rho (k)}{U^{2} + 16(\Lambda - \sin k)^{2}},
\label{liebwumu}
\end{eqnarray}
where $Q_{-}$ and $Q_{+}$ are determined by the conditions
\FL
\begin{equation}
\int_{Q_{-}}^{Q_{+}} dk\, \rho(k) = n, \;\;\;\;\;\;\;\;
\int_{Q_{-}}^{Q_{+}} dk\, k \,\rho(k) = n\Phi_{\rm c}/L,
\label{bound}
\end{equation}
and $\mu$ is normalized to
\begin{equation}
\int_{-\infty}^{\infty} d\Lambda\, \mu(\Lambda) = n/2.
\end{equation}
The ground state energy is then
\begin{equation}
E_0=-2L\int_{Q_{-}}^{Q_{+}} dk\, \cos k \,\rho(k).
\label{enertwist}
\end{equation}

$E_0$ is
independent of $\Phi_{\rm c}$ at $n=1$ in the limit $L \rightarrow
\infty$,
as may be seen
from Eqs.~(\ref{liebwurho})--(\ref{enertwist}):
at half filling, the
pseudomomenta span the entire Brillouin zone $(-\pi,\pi)$, so
the shift in the pseudo-fermi points
$Q_{-}$ and $Q_{+}$ caused by the flux $\Phi_{\rm c}$
merely takes the Brillouin
zone into itself, so that
$\rho(k)$, $\mu(\Lambda)$, and $E_{0}$
are all independent of $\Phi_{\rm c}$ at $n=1$.
To obtain an expression for the ground state energy of the finite
system
suitable for an asymptotic expansion about $L=\infty$,
we express the sum over pseudomomenta in Eq.~(\ref{ener}) as an integral by
means of Eqs.~(\ref{ieqn}), (\ref{theta}), and the Poisson summation
formula, with the result:
\FL
\begin{equation}
E_{0}(\Phi_{\rm c})=-2L\int_{-\pi}^{\pi} d k \cos k \, \rho_{L}(k)
\sum_{m=-\infty}^{\infty} \exp\{i m[L p_{L}(k) - \Phi_{\rm c}
+ 2\pi s]\}.
\label{enerphi}
\end{equation}
In the large-$L$ limit,
the dominant
$\Phi_{\rm c}$-dependence of $E_{0}$ comes from the terms with $m=0,\pm1$.
Moreover, for large
$L$, $p_{L}(k)$ may be approximated by
$p_{\rm c}(k) \equiv {\displaystyle \lim_{L\rightarrow \infty}
p_{L}(k)}$ in the terms with
$m\neq 0$.
$p_{\rm c}(k)$ may be obtained by
integrating the expression for $\rho(k)$ in Ref.~\cite{lwu}, and was
given in Eq.~(\ref{pofk}).
Thus, using Eq.~(\ref{stiff}),
the charge stiffness may be written in the large-$L$ limit as
$D_{\rm c}(L) = D_{\rm c}^{(\delta \rho)} + D_{\rm c}^{(\delta k)}$, where
\FL
\begin{equation}
D_{\rm c}^{(\delta \rho)} = -L^{2} \int_{-\pi}^{\pi} d k \cos k \, \left.
\frac{\delta^{2}
\rho_{L}(k)}{\delta \Phi_{\rm c}^{2}}\right|_{\Phi_{\rm c} = 0},
\label{Ddrho}
\end{equation}
\FL
\begin{equation}
D_{\rm c}^{(\delta k)} = \frac{L}{\pi}\, \int_{-\pi}^{\pi} d k
\sin k \, \sin [L p_{\rm c}(k) + 2\pi s].
\label{Ddk}
\end{equation}
$D_{\rm c}^{(\delta \rho)}$ describes the
flux dependence of the ground state energy
due to the change in the distribution of pseudomomenta as a function of
$\Phi_{\rm c}$,
 while $D_{\rm c}^{(\delta k)}$ describes that due to
a uniform shift of the pseudomomenta, $\delta k_{n}
=\Phi_{\rm c}/L$.  Both vanish as $L \rightarrow \infty$.

The integral in Eq.~(\ref{Ddk})
is dominated by the saddle point
\begin{equation}
k_{0} = \pi + i \sinh^{-1}(U/4),  \label{k0}
\end{equation}
and may be evaluated by the method of stationary phase, yielding
\begin{equation}
D_{\rm c}^{(\delta k)}
= \frac{(-1)^{\frac{L}{2}+1} L^{1/2} (U/4)}{|\pi p_{\rm c}''(k_0)/2|^{1/2}}
\exp[-L/\xi(U)];\;\;\;\;\; L\rightarrow \infty
\label{Ddkass}
\end{equation}
($L$ even), where
\FL
\begin{equation}
1/\xi(U) \equiv -i p_{\rm c}(k_{0}) =
 \frac{4}{U} \int_{1}^{\infty} d y \, \frac{\ln (y +
\sqrt{y^{2} - 1})}{\cosh (2 \pi y/U)}.
\label{xinvA}
\end{equation}

We next consider Eq.~(\ref{Ddrho}) for $D_{\rm c}^{(\delta \rho)}$.
To find $ \delta^{2}
\rho_{L}(k)/\delta \Phi_{\rm c}^{2}|_{\Phi_{\rm c} = 0}$,
we first introduce coupled
integral equations for $\rho_{L}(k)$ and $\mu_{L}(\Lambda)$.  Combining
Eqs.~(\ref{psubl}), (\ref{rho}), and (\ref{liebwurho}), we obtain
\FL
\begin{eqnarray}
\lefteqn{2 \pi [\rho_{L}(k) - \rho(k)]  =  \cos k \int_{-\infty}^{\infty}
d \Lambda \, \frac{8 U[\mu_{L}(\Lambda) -
\mu(\Lambda)]}{U^{2} + 16(\Lambda - \sin k)^{2}}}\nonumber\\
& & \mbox{} +  \cos k \int_{-\infty}^{\infty} d \Lambda \, \frac{8 U}{U^{2}
+ 16(
\Lambda - \sin k)^{2}} \left[ \frac{1}{L} \sum_{\beta = 1}^{M} \delta
(\Lambda -
\Lambda_{\beta}) - \mu_{L}(\Lambda) \right].
\label{rhoint1}
\end{eqnarray}
The Poisson summation formula and Eqs.~(\ref{jeqn}) and (\ref{zed}) can be
used to rewrite Eq.~(\ref{rhoint1}) as
\FL
\begin{eqnarray}
\lefteqn{2 \pi [\rho_{L}(k) - \rho(k)]  =  \cos k \int_{-\infty}^{\infty}
d \Lambda \, \frac{8 U[\mu_{L}(\Lambda) -
\mu(\Lambda)]}{U^{2} + 16(\Lambda - \sin k)^{2}}}\nonumber\\
& & \mbox{} +
\cos k \int_{-\infty}^{\infty} d \Lambda \, \frac{8 U \mu_{L}
(\Lambda)}{U^{2} + 16(\Lambda - \sin k)^{2}}
\sum_{\stackrel{\scriptstyle m=-\infty}{m \neq 0}}^{\infty}
\exp \{i m [L z_{L}(\Lambda) + 2 \pi s']\},
\label{rhoint}
\end{eqnarray}
where $s' = \frac{1}{2} [(M-1)\bmod 2]$.
After similar manipulations, Eqs.~(\ref{zsubl}),
(\ref{sigma}), and
(\ref{liebwumu}) lead to
\FL
\begin{eqnarray}
\lefteqn{
2 \pi [\mu_{L}(\Lambda) - \mu(\Lambda)] =
\int_{-\pi}^{\pi}
d k \, \frac{8 U [\rho_{L}(k) - \rho(k)]}{U^{2} +
16(\Lambda - \sin k)^{2}}
- \int_{-\infty}^{\infty} d \Lambda' \, \frac{4 U [\mu_{L}
(\Lambda') - \mu(\Lambda')]}{U^{2} + 4(\Lambda - \Lambda')^{2}}}
\nonumber \\ & & \mbox{}\;\;\;\;\;\;\;\;\;\;\;\;\;\;\; +
\int_{-\pi}^{\pi}
d k \, \frac{8 U \rho_{L}(k)}{U^{2} + 16(\Lambda - \sin k)^{2}}
\sum_{\stackrel{\scriptstyle m = -\infty}{m \neq 0}}^{\infty}
\exp \{i m [L p_{L}(k) -
\Phi_{\rm c} + 2\pi s]\}
\nonumber \\ & & \mbox{}\;\;\;\;\;\;\;\;\;\;\;\;\;\;\;
- \int_{-\infty}^{\infty} d \Lambda' \, \frac{4 U \mu_{L}
(\Lambda')}{U^{2} + 4(\Lambda - \Lambda')^{2}}
\sum_{\stackrel{\scriptstyle m = -\infty}{m \neq 0}}^{\infty}
\exp \{i m [L z_{L}(\Lambda')
 + 2\pi s' ] \}.
\label{sigmaint}
\end{eqnarray}
Differentiating Eqs.~(\ref{rhoint}) and (\ref{sigmaint}) twice with
respect to $\Phi_{\rm c}$,
and keeping only the leading order terms, we obtain the
following coupled integral equations for $\delta^{2}
\rho_{L}(k)/\delta \Phi_{\rm c}^{2}|_{\Phi_{\rm c} = 0}$ and
$\delta^{2} \mu_{L}(\Lambda)/
\delta \Phi_{\rm c}^{2}|_{\Phi_{\rm c} = 0}$,
valid in the large-$L$ limit:
\FL
\begin{eqnarray}
2 \pi \left.\frac{\delta^{2}
\rho_{L}(k)}{\delta \Phi_{\rm c}^{2}}\right|_{\Phi_{\rm c} = 0} & = & \cos k
\int_{-\infty}^{\infty} d \Lambda \, \frac{8 U}{U^{2} + 16(\Lambda -
\sin k)^{2}}\left.
\frac{\delta^{2} \mu_{L}(\Lambda)}{\delta
\Phi_{\rm c}^{2}}\right|_{\Phi_{\rm c} = 0},
\rule[-1cm]{0cm}{1cm}
\label{drhodphi} \\
2 \pi
\left.\frac{\delta^{2} \mu_{L}(\Lambda)}{\delta \Phi_{\rm c}^{2}}\right|_{
\Phi_{\rm c} = 0}
& + & \int_{-\infty}^{\infty} d \Lambda' \, \frac{4U}{U^{2} + 4(\Lambda -
\Lambda')^{2}}
\left.\frac{\delta^{2} \mu_{L}(\Lambda')}{\delta \Phi_{\rm c}^{2}}\right|_{
\Phi_{\rm c} = 0}
\nonumber \\
& & \mbox{} =
\int_{-\pi}^{\pi} d k \, \frac{8U}{U^{2} + 16(\Lambda - \sin k)^{2}}
\left.\frac{\delta^{2}\rho_{L}(k)}{\delta
\Phi_{\rm c}^{2}}\right|_{\Phi_{\rm c} = 0}
\nonumber \\ & & \mbox{} -  \int_{-\pi}^{\pi}
d k \, \frac{16 U \; \rho_{\infty}(k)}{U^{2} + 16(\Lambda - \sin k)^{2}}
\cos [L p_{\rm c}(k) + 2\pi s].
\label{dsigdphi}
\end{eqnarray}
These equations can be solved by Fourier transforms, with the result
\FL
\begin{equation}
\left.\frac{\delta^{2} \mu_{L}(\Lambda)}{\delta
\Phi_{\rm c}^{2}}\right|_{\Phi_{\rm c}
= 0}
 =  -\int_{-\pi}^{\pi}\frac{dk}{2\pi} \rho(k) \cos[Lp_{\rm c}(k) +2\pi s]
\int_{-\infty}^{\infty} \frac{\exp[i\omega(\Lambda-\sin k)]\, d\omega}{2
\cosh (\omega U/4)},
\label{dmudphians}
\end{equation}
\FL
\begin{equation}
\left.\frac{\delta^{2}
\rho_{L}(k)}{\delta \Phi_{\rm c}^{2}}\right|_{\Phi_{\rm c} = 0}
 =  - \int_{-\pi}^{\pi} \frac{dk'}{2\pi} \rho(k')
 \cos [Lp_{\rm c}(k') + 2\pi s]
\frac{\partial \Theta(k,k')}{\partial k},
\label{drhodphians}
\end{equation}
where $\Theta(k,k')$ was defined in Eq.~(\ref{gamma}).  Inserting
Eq.~(\ref{drhodphians}) into Eq.~(\ref{Ddrho}), and evaluating the integral
over $k'$ by the method of stationary phase in the large-$L$ limit, we obtain
\FL
\begin{eqnarray}
D_{\rm c}^{(\delta \rho)} & = & \frac{(-1)^{\frac{L}{2}} L^{1/2} [1+(4/U)^2
]^{1/2} \exp[-L/\xi(U)]}{|\pi p_{\rm c}''(k_0)/2|^{1/2}}\nonumber\\
& & \mbox{} \times \int_{0}^{\infty} dx \, e^{-x} \tanh (x) J_1 (4x/U);
\;\;\;L\rightarrow\infty
\label{Ddrhoans}
\end{eqnarray}
($L$ even).
The sign of $D_{\rm c}^{(\delta
\rho)}$ is opposite that of $D_{\rm c}^{(\delta k)}$.
As noted in Ref.~\cite{ours}, $D_{\rm c}^{
(\delta \rho)}$ is negligible compared to
$D_{\rm c}^{(\delta k)}$ in the large-$U$
limit.  In the small-$U$ limit they are comparable in magnitude, but
$|D_{\rm c}^{(\delta k)}| > |D_{\rm c}^{(\delta \rho)}|\;\;\forall\; U$.
Combining Eqs.~(\ref{Ddkass}) and (\ref{Ddrhoans}) yields
Eq.~(\ref{stiffans}), with
\FL
\begin{equation}
D(U)= \frac{(U/4)^2 - [1+
(U/4)^{2}]^{1/2} \int_{0}^{\infty} dx \, e^{-x} \tanh
(x) J_1 (4x/U)}{\left( \frac{\pi}{2} \int_{0}^{\infty}
dx \, e^{-x} J_0 (4x/U) \{(U/4)^2 + [1 + (U/4)^2]
x \tanh (x)\}\right)^{1/2}}.
\label{eq.coeff}
\end{equation}

By applying arguments similar to those above directly to Eq.~(\ref{enerphi}),
one can readily derive Eq.~(\ref{eperiodic}).

\appendix{CALCULATION OF $D_{\rm c}$ NEAR HALF FILLING}
\label{appB}

Here we use a technique due to Haldane \cite{haldane3} to calculate the charge
stiffness near half filling, verifying the term linear in $\delta$ in
Eq.~(\ref{stiffholon}).
Eliminating $\mu(\Lambda)$ from Eqs.~(\ref{liebwurho}) and (\ref{liebwumu})
yields an integral equation for $\rho(k)$ alone \cite{coll}:
\begin{equation}
2\pi \rho(k) = 1 + \int_{Q_{-}}^{Q_{+}} dk'\,
\frac{\partial \Theta(k,k')}{\partial k} \rho(k'),
\label{rhotwist}
\end{equation}
where
$\Theta(k,k')$ was defined in
Eq.~(\ref{gamma}).
We note that
Eqs.~(\ref{bound}), (\ref{enertwist}), and (\ref{rhotwist}), which determine
the dependence of the ground state energy on $\Phi_{\rm c}$ in the limit $L
\rightarrow \infty$, are
identical in form to those considered by Haldane in his treatment of the
spinless 1D quantum fluid \cite{haldane3}.  The result for the charge
stiffness obtained in Ref.~\cite{haldane3} is therefore applicable here,
and is
\begin{equation}
D_{\rm c} = \frac{e^{2\phi} v_{\rm c}}{2 \pi},
\label{vj}
\end{equation}
where $e^{\phi}=1-\omega(Q)-\omega(-Q)$ is a positive number which
parameterizes the electron-electron interactions \cite{haldanellt}, and
\begin{equation}
v_{\rm c} = \frac{1}{2\pi \rho(Q)}
\left(\varepsilon'(Q) - \int_{-Q}^{Q} dk \, \varepsilon'(k) \tau(k)\right)
\label{vcharge}
\end{equation}
is the charge velocity.
Here $\varepsilon(k) =-2t\cos k$, and $\omega(k)$ and $\tau(k)$ are solutions
of the following integral equations:
\FL
\begin{equation}
2 \pi \omega(k) = \Theta(k,Q) -
\int_{-Q}^{Q} dk'\, \frac{\partial \Theta(k,k')
}{\partial k'} \omega(k'),
\label{omega}
\end{equation}
\FL
\begin{equation}
2 \pi \tau(k) = \frac{\partial \Theta(k,Q)}{\partial Q} -
\int_{-Q}^{Q} dk'\, \frac{\partial \Theta(k,k')}{\partial k'} \tau(k').
\label{mu}
\end{equation}

Using Eq.~(\ref{gamma}) for $\Theta(k,k')$
and the fact that $Q=\pi$ at half filling, we see that
$v_{\rm c}$ vanishes at half filling because $\varepsilon'(\pi) =0$ and
$\tau(-k) = \tau(k)$ when $Q=\pi$.  Also, $\omega(\pi)=\omega(-\pi)=0$, so
$e^{\phi}=1$ at half filling.
Thus, for a small doping $\delta$ away from half filling
\FL
\begin{equation}
D_{\rm c} \simeq
-\delta\left.\frac{\partial D_{\rm c}}{\partial n}\right|_{n=1^-}=
-\delta\left.\frac{\partial D_{\rm c}}{\partial Q}\right|_{Q=\pi}
\left.\frac{\partial Q}{\partial n}\right|_{n=1^-}
\label{Ddope}
\end{equation}
 From Ref.~\cite{haldane3} $\partial Q/ \partial n =
[1-\omega(Q)+\omega(-Q)]/2\rho(Q)$,
so $\left.(\partial Q/ \partial n)\right|_{n=1^-} = 1/2\rho(\pi)$.  After some
manipulation, Eq.~(\ref{Ddope}) becomes
\FL
\begin{equation}
D_{\rm c} \simeq  \frac{t\delta}{[2\pi \rho(\pi)]^{2}}
\left( 1 + \frac{1}{2\pi}
\int_{-\pi}^{\pi} \sin k \left.\frac{\partial^{2}
\Theta(k,Q)}{\partial Q^{2}}\right|_{Q=\pi} dk\right).
\label{Ddope2}
\end{equation}
Inserting the result for $\rho(\pi)$ from Ref.~\cite{lwu}
into Eq.~(\ref{Ddope2}) and
using Eq.~(\ref{gamma}), we obtain \cite{kaw2}
\FL
\begin{equation}
D_{\rm c} \simeq t \delta
\left.\left(1-2\int_{0}^{\infty} \frac{\omega J_{1}(\omega)\,d\omega}{1 +
\exp(\omega U/2t)}\right)\right/\left(1-2\int_{0}^{\infty}
\frac{J_{0}(\omega)\,d\omega}{1 +
\exp(\omega U/2t)}\right)^{\textstyle 2}.
\label{DdU}
\end{equation}
Note that the above expression for $D_{\rm c}$
may be rewritten using Eqs.~(\ref{mdef}), (\ref{pfirst}), and
(\ref{esecond}) as $D_{\rm c} \simeq \delta/|2m^{\ast}|$,
thus confirming the term linear in $\delta$ in Eq.~(\ref{stiffholon}).
Note also that $v_{\rm c} \simeq \pi \delta/|m^{\ast}|$ for $\delta \ll 1$, in
agreement with Eq.~(\ref{chargevel}).

For completeness, we give below the explicit expressions for
$p_{\rm c}'(\pi)$, $\varepsilon_{\rm
c}''(\pi)$, $p_{\rm c}^{(3)}(\pi)$, and $\varepsilon_{\rm c}^{(4)}(\pi)$:
\begin{equation}
p_{\rm c}'(\pi) =
1-2\int_{0}^{\infty} \frac{J_{0}(\omega)
\,d\omega}{1 +
\exp(\omega U/2t)},
\label{pfirst}
\end{equation}
\begin{equation}
\varepsilon_{\rm c}''(\pi) = - 2t
\left(1-2\int_{0}^{\infty} \frac{\omega J_{1}(\omega)\,d\omega}{1 +
\exp(\omega U/2t)}\right),
\label{esecond}
\end{equation}
\begin{equation}
p_{\rm c}^{(3)}(\pi) =
2\int_{0}^{\infty}
\frac{(1 + \omega^2) J_{0}(\omega) \,d\omega}{1 + \exp(\omega U/2t)},
\label{pthird}
\end{equation}
\begin{equation}
\varepsilon_{\rm c}^{(4)}(\pi) =  2t
\left(1-2\int_{0}^{\infty} \frac{(\omega^3 + 4 \omega)
J_{1}(\omega)\,d\omega}{1 + \exp(\omega U/2t)}\right).
\label{efourth}
\end{equation}

Using the Laplace transforms of $J_0$ and $J_1$, and the Sommerfeld-Watson
transformation,
Eq.~(\ref{DdU}) may be rewritten in a form more suitable for an asymptotic
expansion about $U=0$ as follows:
\FL
\begin{equation}
D_{\rm c} \simeq t \delta
\left.\left(\frac{\pi^{2}t}{U} \int_{1}^{\infty} dy\,
\frac{(y^{2}
-1)^{1/2}
[2\coth^{2}(2\pi t y/U) -1]}{\sinh(2\pi t y/U)}\right)\right/
\left(\int_{1}^{\infty} dy\,
\frac{(y^{2}-1)^{-1/2}}{\sinh(2\pi t y/U)}\right)^{\textstyle 2}.
\label{Ddu}
\end{equation}


\figure{Ground state energy of a half-filled Hubbard ring versus
magnetic flux $(\hbar c/e)\Phi_{\rm c}$ from $\Phi_{\rm c}=0$ to $2\pi$.
Solid curves:  $E_0(\Phi_{\rm c})+ N|U|/2$ for $U=-4$;
dashed curves:  $E_0(\Phi_{\rm c})$
for $U=4$.  (a) $N=L=6\approx 1.5\xi$; (b) $N=L=28\approx 7\xi$, where
$\xi(4) \simeq 4.06$ is the correlation length at half filling for
$|U|=4$.
\label{fluxquantum}}

\figure{
Plots of $\pi D_{\rm c}$ versus $\xi/L$
for (a) $\delta=0$, (b) $\delta=1/L$, (c) $\delta=2/L$, and (d) $\delta=4/L$
for systems with $N=60$ (diamonds),
$N=80$ (crosses), and $N=100$ (squares), and $0.5 \leq U \leq 2$,
illustrating the scaling law,
Eq.~(\ref{hyperscale2}).
Note the different vertical scales.
(e) For comparison, (b) is replotted with
$\xi$ as the abscissa.
\label{fig.hyperscaling}}

\figure{Total spectral weight $\pi N_{{\rm tot}}$ (squares) and Drude weight
$\pi D_{\rm c}$ (triangles) as a function of system size $L$ for $U=6$ and
$\delta=0.2$.  The solid curves indicate $L^{-2}$ behavior
extrapolated to small $L$.
\label{finitesize}}

\figure{Plots of (a) $A(\delta,U)$ and (b) $B(\delta,U)$ as a function of $U$
for $\delta=0.05$, 0.1, 0.2, and 0.4 (top to bottom),
where $D_{\rm c}(L)=D_{\rm c}(\infty)
[1+A(\delta,U)/L^2 +\cdots]$ and $N_{{\rm tot}}(L)=N_{{\rm tot}}(\infty)[1+
B(\delta,U)/L^2 +\cdots]$.  Note the different vertical scales.
\label{largesize}}

\figure{A plot of the function
$f_+\equiv [A(\delta,U)-\pi^2/6]
/\xi(U)^2$
versus $\xi(U)\delta$.  Solid curve:  $\delta^{-1}=41$; $1.8\leq U\leq 4.7$.
Triangles:  $U=3$; $\delta^{-1}=10$, 15, 20, 30, 40, 60.  Dotted curve:
$f_+(x)=
\pi^2/2-\pi^4 x^2$, the analytic result in the small-$x$
limit.\label{scaling}}


\begin{references}
\bibitem{triplet} The minimum energy eigenvalue for fixed $N$, $L$, and $U$
is  always obtained in the singlet state and may occur at either
$\Phi_{\uparrow}=\Phi_{\downarrow}=0$
or $\Phi_{\uparrow}=\Phi_{\downarrow}=\pi$.  However, when $N\neq L$ and
$U>0$ there is a level crossing with the triplet state (see
Refs.~\cite{ogatashiba,liebmattis}).
\bibitem{xinv2} To facilitate an asymptotic expansion of $\xi(U)$
about $U=\infty$, one can rewrite Eq.~(\ref{xinv}) as
$1/\xi(U)  =
\sinh^{-1}(U/4) - \int_{0}^{\infty}
d x\, x^{-1} e^{-x} \tanh x\, J_{0}(4x/U)$.
\bibitem{bandw} This may be seen most easily in Brillouin-Wigner
perturbation theory, where the energy denominators need only be
kept to ${\cal O}(U)$ to obtain the leading order expression for $G$.
\bibitem{refschulz} This result was obtained previously for the case $L
= \infty$ in Ref.~\cite{schulzcond}.
\bibitem{boundcond} The finite-size corrections are more complicated for
other choices of boundary conditions, and may be either positive
or negative, depending on $N$, $L$, and $U$.
\bibitem{Lminus2} In the insulating phase, the finite-size corrections
to $D_{\rm c}$ are of course exponential in $L$,
but those to $N_{{\rm tot}}$ are
still ${\cal O}(L^{-2})$ (see Ref.~\cite{ours}).
\bibitem{exception} The case $n=1$ is an exception, where the
finite-size corrections to $D_{\rm c}$ near the critical point
have opposite signs and vastly
different magnitudes for periodic and antiperiodic boundary conditions
when $L<\xi$, as discussed in
Refs.~\cite{ours,fye}.
\bibitem{phaseshift} Note that $\Theta(k,k')$ is the same effective two-body
scattering phase shift which appears in the Bethe ansatz integral
equation for the ground state distribution of pseudomomenta in the
infinite system, Eq.~(\ref{rhotwist}).
\bibitem{Drudefree} We remark that Eq.~(\ref{Drude}) remains valid when
terms of ${\cal O}(\delta^2)$ are included, provided we replace
$1/m^{\ast}$ by
$d^2 \varepsilon_{\rm c}/d p^2|_{p=\pi(1-\delta)}$
for $n < 1$ and by
$d^2 \varepsilon_{\rm c}^{\rm UHB}/d p^2|_{p=\pi(1-\delta)}$
for $n > 1$.
The relation $2\partial D_{\rm c}/\partial n =
\varepsilon_{\rm c}''(p_F)$ is generically true for noninteracting
fermions in a 1D energy band $\varepsilon_{\rm c}(p)$, provided there
are only two Fermi points.
\bibitem{impurities} The experimental situation is complicated by the
fact that the impurity potential introduced by the dopants localizes the
charge carriers at small dopant concentrations.
Thus the doping regime which Eq.~(\ref{thermopower}) might be expected
to describe may not be accessible experimentally.
At larger dopings, the thermopower of the 1D Hubbard model has
a contribution from
the entropy carried by the spin excitations, which leads to
an unusual temperature dependence (see Ref.~\cite{myprl}).
\bibitem{coll} B.~S.~Shastry, private communication.
Equation (\ref{rhotwist}) differs from that derived by H.~Shiba, Phys.
Rev. B {\bf 6}, 930 (1972), only by the shift of the
pseudo-fermi points from $-Q$, $Q$ to $Q_-$, $Q_+$, and may be derived by the
same method.
\end{references}
\end{document}